\documentclass[11pt]{elsarticle}

\usepackage{lineno}
\usepackage{caption}
\usepackage{subcaption}
\usepackage{multirow}
\usepackage{multicol}
\usepackage[table]{xcolor}

\usepackage{listings}

\usepackage{graphicx}
\usepackage{xcolor}
\def\BibTeX{{\rm B\kern-.05em{\sc i\kern-.025em b}\kern-.08em
    T\kern-.1667em\lower.7ex\hbox{E}\kern-.125emX}}

\usepackage{CJKutf8}
\def\BibTeX{{\rm B\kern-.05em{\sc i\kern-.025em b}\kern-.08em
    T\kern-.1667em\lower.7ex\hbox{E}\kern-.125emX}}

\newcommand{\etal}{\textit{et al}. }

\usepackage{amsmath,amssymb,amsfonts}

\usepackage{glossaries}
\usepackage{verbatim}
\usepackage{soul}
\usepackage{xurl}

\newacronym{api}{API}{Application Programming Interface}
\newacronym{ai}{AI}{Artificial Intelligence}
\newacronym{cpss}{CPS}{Cyber-Physical Systems}
\newacronym{cps}{CPS}{Cyber-Physical System}
\newacronym{cti}{CTI}{Cyber Threat Intelligence}
\newacronym{qd}{QD}{Quantum Defense}
\newacronym{dos}{DoS}{Denial of Service}
\newacronym{ep}{EP}{Exceedance Probability}
\newacronym{scada}{SCADA}{Supervisory Control and Data Acquisition}
\newacronym{ict}{ICT}{Information and Communications Technology}
\newacronym{llm}{LLM}{Large Language Model}
\newacronym{lti}{LTI}{Linear Time Invariant}
\newacronym{lqg}{LQG}{Linear Quadratic Gaussian}
\newacronym{lqr}{LQR}{Linear Quadratic Regulator}
\newacronym{lpa}{LPA}{Label Propagation Algorithm}
\newacronym{mdp}{MDP}{Markov Decision Process}
\newacronym{cdf}{CDF}{Cumulative Distribution Function}
\newacronym{ml}{ML}{Machine Learning}
\newacronym{mtd}{MTD}{Moving Target Defense}
\newacronym{ncss}{NCSs}{Networked-Control Systems}
\newacronym{ncs}{NCS}{Networked-Control System}
\newacronym{pml}{PML}{Probable Maximum Loss}
\newacronym{rl}{RL}{Reinforcement Learning}
\newacronym{sdn}{SDN}{Software Defined Networking}
\newacronym{sutd}{SUTD}{Singapore University of Technology and Design}
\newacronym{swat}{SWaT}{Secure Water Treatment}
\newacronym{qml}{QML}{Quantum Machine Learning}
\newacronym{plc}{PLC}{Programmable-Logic Controller}
\newacronym{uf}{UF}{Ultrafiltration}
\newacronym{ro}{RO}{Reverse Osmosis}
\newacronym{tmp}{TMP}{Transmembrane Pressure}
\newacronym{ic}{IC}{Integrated Circuit}
\newacronym{or}{OR}{Organizational Resilience}
\newacronym{iot}{IoT}{Internet of Things}
\newacronym{fis}{FIS}{Fuzzy Inference System} 
\newacronym{frb}{FRB}{Fuzzy Rule Base} 
\newacronym{iira}{IIRA}{The Industrial Internet Reference Architecture}
\newacronym{rami}{RAMI 4.0}{Reference Architectural Model Industrie 4.0}
\newacronym{pfs}{PFS}{Secure Ports of the Future}
\newacronym{gdpr}{GDPR}{General Data Protection Regulation}
\newacronym{it}{IT}{Information Technology}
\newacronym{ot}{OT}{Operational Technology}
\newacronym{std}{STD}{Standard Deviation}
\newacronym{dynamo}{DYNAMO}{Dynamic Resilience Assessment Method including a combined Business Continuity Management and Cyber Threat Intelligence solution for Critical Sectors}
\newacronym{uv}{UV}{Ultraviolet} 
\newacronym{wcc}{WCC}{Weakly Connected Component} 
\newacronym{fftx}{FFTx}{Fiber-to-the-x}
\newacronym{ttp}{TTP}{Tactics, Techniques, and Procedures}
\newacronym{apt}{APT}{Advanced Persistent Threat}
\newacronym{ics}{ICS}{Industrial Control System}
\newacronym{europol}{Europol}{European Union Agency for Law Enforcement Cooperation}
\newacronym{raas}{RaaS}{Ransomware as a Service}

\usepackage{soul}

\usepackage{hyperref} %<--- Load after everything else
\hypersetup{
  colorlinks=true,
  linkcolor=blue,
  urlcolor=blue,
  citecolor=blue,
}

\journal{~~}

\begin{document}

\begin{frontmatter}

\title{Graph Analytics for Cyber-Physical\\ System Resilience Quantification}

\author[inst1,inst3]{Romain Dagnas}
\affiliation[inst1]{organization={IRT SystemX},%Department and Organization
            %addressline={}, 
            city={Palaiseau},
            postcode={91120}, 
            country={France}}

\author[inst2]{Michel Barbeau}
\affiliation[inst2]{organization={Carleton University},%Department and Organization
            %addressline={}, 
            city={Ottawa},
            postcode={ON K1S 5B6}, 
            country={Canada}}
            
\author[inst3]{Joaquin Garcia-Alfaro}
\affiliation[inst3]{organization={SAMOVAR, Télécom SudParis, Institut Polytechnique de Paris},%Department and Organization
            %addressline={}, 
            city={Palaiseau},
            postcode={91120}, 
            country={France}}

\author[inst1]{Reda Yaich}

\begin{abstract}
Critical infrastructures integrate a wide range of smart technologies and become highly connected to the cyber world. This is especially true for Cyber-Physical Systems (CPSs), which integrate hardware and software components. Despite the advantages of smart infrastructures, they remain vulnerable to cyberattacks. This work focuses on the cyber resilience of CPSs. We propose a methodology based on knowledge graph modeling and graph analytics to quantify the resilience potential of complex systems by using a multilayered model based on knowledge graphs. Our methodology also allows us to identify critical points. These critical points are components or functions of an architecture that can generate critical failures if attacked. Thus, identifying them can help enhance resilience and avoid cascading effects. We use the SWaT\texttrademark~(Secure Water Treatment) testbed as a use case to achieve this objective. This system mimics the actual behavior of a water treatment station in Singapore. We model three resilient designs of SWaT according to our multilayered model. We conduct a resilience assessment based on three relevant metrics used in graph analytics. We compare the results obtained with each metric and discuss their accuracy in identifying critical points. We perform an experimentation analysis based on the knowledge gained by a cyber adversary about the system architecture. We show that the most resilient SWaT design has the necessary potential to bounce back and absorb the attacks. We discuss our results and conclude this work by providing further research axes.

\end{abstract}


\end{frontmatter}

%\linenumbers

\section{Introduction}
\label{sec:introduction}

\noindent During recent years, incidents such as the WannaCry ransomware~\cite{wannacry2017}, the SolarWinds software provider attack~\cite{solarwinds2021}, and more recently, the Lockbit group  activities~\cite{lockbit} have highlighted the vulnerabilities of complex systems, i.e., \glspl*{cps} facing cyber adversaries as discussed by Riggs~\etal \cite{riggs2023impact}. Every strategic sector is affected, including maritime, mobility, and energy. Protecting critical infrastructures is paramount, especially in our era, where cyber attacks can generate cascading effects, leading to significant losses.

The resilience concept has gained attention in the \glspl*{cps} research community. Industrial entities gradually understand the importance of designing resilient systems and enhancing the resilience of existing architectures. 
The notion of resilience refers to the ability of a system to continue to operate and complete a mission, even if an adversarial event occurs. This adversarial event can be natural or intentional. Kott and Linkov define resilience as \textit{a system’s ability to recover or regenerate its performance after an unexpected impact produces a degradation of its performance} \cite{linkov2013resilience}. Resilience was initially applied in ecology by Holling \cite{Holling1973ResilienceAS} to quantify a population's ability to recover from changes. 
As explained by Hosseini~\etal~\cite{hosseini2016resi}, resilience is now applied in many other fields, for example, psychology, economy, engineering, computer science, and cybersecurity. Linkov and Kott~\cite{kott2021resilience}  mention that \textit{to improve the cyber resilience of a system, you have to measure it}.

Measurement of the resilience of a \gls*{cps} implies using and creating metrics based on certain system properties.
\glspl*{cps} and especially critical infrastructures increasingly connect and include many components. 
Their architectures become increasingly complex (e.g., architecture design, human workflows, and operating environment). Due to this complexity, building accurate models of such systems is not easy. 
The less accurate the model, the lower the accuracy of the assessment resulting from using a metric. Furthermore, the more complex the architecture, the more prone it is to cascading effects.
In cybersecurity, we consider the case of intentional cyber disruptive actions perpetrated to generate impacts of a high magnitude.

In the \gls*{cps} context, we define cascading effects related to \glspl*{cps} as unexpected and unintentional sequences of events that start with a disruptive action and spread through an architecture in which the subsystems have dependencies, generating significant losses.

\medskip

\textbf{Motivation.} Due to their complexity, critical infrastructure and complex \glspl*{cps} are subjected to cascading effects, i.e., attacks targeting one specific point that have unexpected repercussions on other functions or components of system architecture. These cascading effects are difficult to anticipate. They can have devastating consequences. The challenge of improving the resilience of complex \glspl*{cps} is akin to building barriers that make attacks very difficult to perpetrate for cyber adversaries and to stop the propagation of their effects. 

\medskip

\textbf{Contribution.} This work is an extended version of our previous work dedicated to quantifying the resilience of multilayered \glspl*{cps}~\cite{dagnas2024multi}\footnote{This a revised and extended version of a paper~\cite{dagnas2024multi} that appeared in the proceedings of the 23rd IFIP International Conference on Networking (IFIP NETWORKING), Thessaloniki, Greece, June 12-15, 2024.}, in which we introduced a multilayered model based on knowledge graphs to perform a quantitative resilience assessment of the pumping subsystem of the SWaT (Secure Water Treatment) testbed. 

\medskip

The contributions in this new work are fourfold: 

\begin{enumerate}

\item We build three designs of the complete SWaT architecture with our multilayered model to conduct a resilience assessment analysis. 

\item We consider three graph analytics metrics that are relevant for resilience assessment. We apply these metrics to the three designs of SWaT to identify critical points. We compare the results obtained and we discuss their precision in identifying critical points.

\item We analyze how a cyber attacker can identify these critical points. The adversary can train a LLM (Large Language Model), generating a graph of the SWaT architecture and attack critical points to generate cascading effects. Building on this experimentation, we determine if the most resilient SWaT design is sufficiently resilient to protect critical points. 

\item Finally, we discuss the result obtained and present countermeasures and strategies that improve resilience.
\end{enumerate}

\medskip

In Section~\ref{sec:background}, we provide background material,
survey related work,
and describe our methodology for quantifying and improving resilience potential. Section~\ref{sec:swatra} provides experimental evaluation results using three designs of SWaT. Section~\ref{sec:experimentation} is dedicated to experimentation based on a cyber adversary trying to find critical points. We also discuss the results obtained. Section~\ref{sec:conclusion} provides the conclusion and perspectives for future work.

\section{Background and Motivation}
\label{sec:background}

\noindent In this section, we review relevant material on the resilience of \glspl*{cps}.

\subsection{Resilience of Cyber-Physical Systems}

\noindent In our era of digitization, increased competitiveness implies a transformation of critical infrastructure architectures. These digitized complex systems must remain competitive and, at the same time, must ensure the completion of specific missions. They become more connected and smarter. Despite its advantages, this digitization comes with new threats that cyber adversaries can exploit. Quantifying and enhancing the resilience of complex systems is a challenge due to this complexity. However, strengthening the resilience of critical infrastructures is of paramount importance. 
Despite protection and security measures, systems continue to face new threats. Cyber adversaries attempting to target critical infrastructures exploit hardware and software vulnerabilities and social engineering techniques.

The case of LockBit is very interesting. It was the world's most active ransomware group from until its dismantling in February 2024 with a task force created by \gls*{europol}. Lockbit was a \gls*{raas}. It operated under a business model with future investments in operations, public relations and recruitment processes. Lockbit was one of the most active malware in the world, affecting hospitals, city halls, and companies of all sizes \cite{lockbitfall}.

More recently, critical infrastructures have been exposed to a new family of cyber attacks. Web-based technologies are widely used by operators responsible for supervising these complex systems. There exists a way to hijack \glspl*{plc}, which includes embedded webservers, making them accessible from cyberspace. According to security experts, adversaries can gain full access to the system by exploiting these architectures. Researchers refer to a surface of attack that has been neglected for many years \cite{georgianethijack}, which is very alarming.

The notion of resilience in cybersecurity aims to protect \glspl*{cps} from adversarial events. In resilience, the \textit{zero} risk does not exist. We assume that cybercriminals can bypass security measures. Based on this assumption, the main challenge is to ensure that complex \glspl*{cps} have the resilience potential necessary to absorb attacks and continue to operate. 

Cledel~\etal presented various resilience definitions existing in all fields \cite{cledel2020resilience}. The first definition of resilience was introduced by Holling in the field of ecology in 1973, as \textit{the capacity of a system to move from a stability domain to another} \cite{Holling1973ResilienceAS}. Resilience has reached other fields, such as psychology, where Southwick~\etal define it as \textit{the process of adapting well in the face of adversity, trauma, tragedy, threats, or even significant sources of stress} \cite{southwick2014resilience}. More recently, resilience gained interest in engineering, where Francis~\etal define it as \textit{a system property to endure undesired events to ensure the continuity of normal system function} \cite{francis2014metric}.
In computer science and cybersecurity, resilience is described by Linkov~\etal as \textit{the system’s ability to recover or regenerate its performance after an unexpected impact produces a degradation of its performance} \cite{linkov2013resilience}.

This definition highlights two important pillars of cyber resilience. The first is monitorability. We must be able to conduct attack detection strategies based on observing the system's behavior. Without enough monitorability, we cannot detect malicious actions. The second pillar is steerability. The definition of resilience describes the ability to return to a stable state. This can be done through actuators. Once we detect a malicious action, the system must be able to act to absorb the attack and continue to operate. 

\subsection{Cyber Resilience Analytics}
\label{sec:graph_analytics}

\noindent This section presents the necessary material for considering multilayered approach based on knowledge graph modeling for resilience quantification and enhancement purposes.

In our previous work, we have presented a way to quantify the resilience potential of a system with the $(k,\ell)$-resilience property (giving an estimation of the steerability degree $k$ and the monitorability degree $\ell$ of a \gls*{cps}) \cite{barbeau2020metrics,barbeau2021resilience}. Indeed, increasing a system's resilience implies monitoring it (by the bias of sensors) and controlling it (by the bias of actuators) to bring it back to its original state in case of an attack. We have also presented an approach using the spectral radius metric to quantify the resilience of a system modeled with a graph structure \cite{dagnas2023exploring}. We must highlight that we consider \glspl*{ncs}. In other fields, such as biology, self-healing systems can restore themselves. The resilience of \gls*{cps} is similar from the point of view of recovery. Resilient systems can recover from disruptions due to their monitorability and steerability capabilities.

Resilience assessment must consider the system from a perspective that considers relationships between components and entities. Knowledge graphs are a way to model complex systems with diversified links and relationships.

\subsubsection{Knowledge Graphs}

\noindent Ehrlinger and Woess reviewed several definitions of the concept \emph{knowledge graph} that we can find in the literature~\cite{ehrlinger2016towards}. Early definitions highlight that knowledge graphs use ontologies to acquire information and apply reasoning mechanisms to derive new knowledge about this information. 
Other definitions go further and differ between fields. For developers, knowledge graphs are similar to a database with which we interact with the bias of an \glspl*{api}. For data scientists, it corresponds to an augmented feature store for connected data, where we can compute and access structural features for \gls*{ml}. For data engineers, it is similar to a data store where we can integrate data from different sources. As explained by Barrasa, it is a database linked to a front-end interface for other fields, with which we can communicate with \cite{barrasa2023knowledgegraphdef}.

In the cyber-resilience field, we consider knowledge graphs for their ability to model various entities and their relationships. Knowledge graphs can be used to model the knowledge acquired by an adversary to carry out high-impact attacks. Defenders can also use knowledge graphs to anticipate cascading effects of attacks perpetrated at critical points. We must highlight that knowledge graphs are also interesting for building remediation graphs to provide specific actions to avoid cascading effects and significant losses. When considering powerful adversaries aiming to disrupt a system by injecting data, knowledge graphs also have an interest in truth discovery to correct these compromised values \cite{li2016survey,wang2024survey}.

\subsubsection{Multilayered Architecture and Resilience}
\label{sec:multi_layer}

\noindent Multilayered strategies allow one to consider the different levels of a system independently and analyze each of these layers. Our objective is to perform a resilience analysis on each layer of a multilayered model to ensure that the resilience potential of all these layers is consistent with the others. Critical infrastructures are complex systems.
Conducting a resilience analysis on such an architecture is a difficult task. 
However, how can we ensure that a proven effective resilience countermeasure does not negatively impact the resilience of another layer? We must remember that increasing a system's resilience potential can also increase the attack surface. In fact, in previous work \cite{barbeau2021resilience}, we have shown that increasing the monitorability and steerability of a \gls*{cps} increases its resilience. This implies a diverse architecture with monitoring, sensors, and steerability components, i.e., pumps and valves, for water treatment purposes. However, our analysis also shows that having more components connected to cyberspace can increase the attack surface. Thus, a fine balance must be achieved between increasing resilience potential and mitigating security risk. Resilience analysis and risk analysis must be conducted in concert.

\begin{figure}[!ht]
\centering
\includegraphics[width=0.49\textwidth]{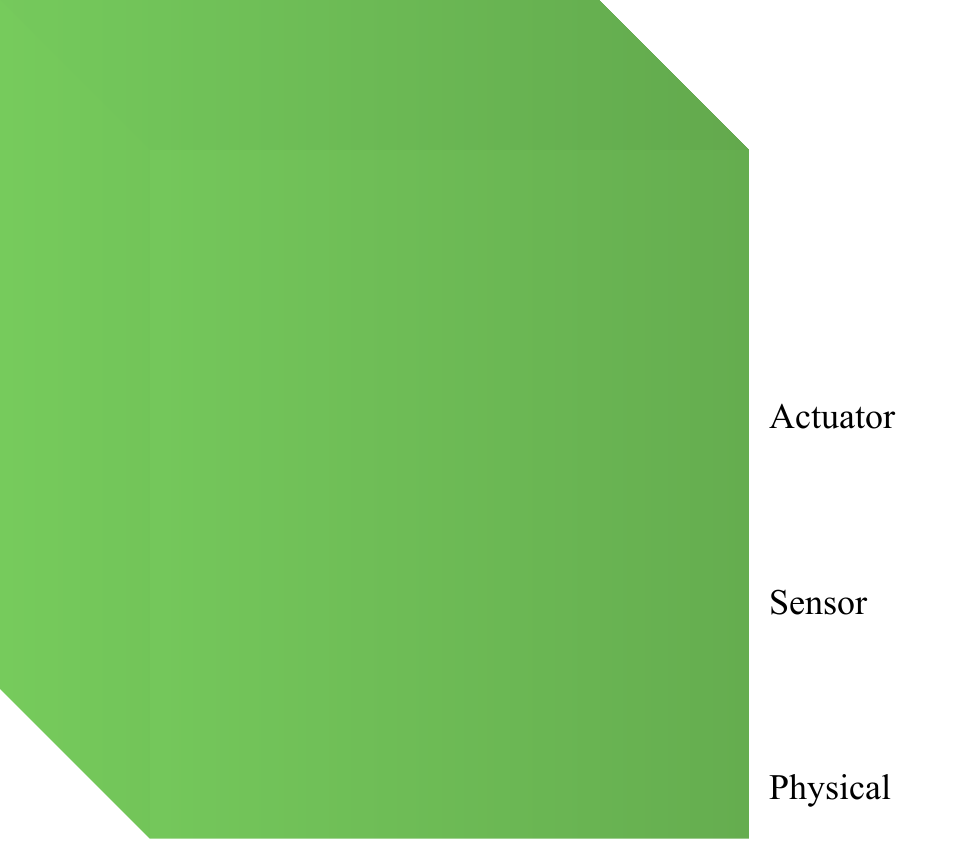}
\includegraphics[width=0.49\textwidth]{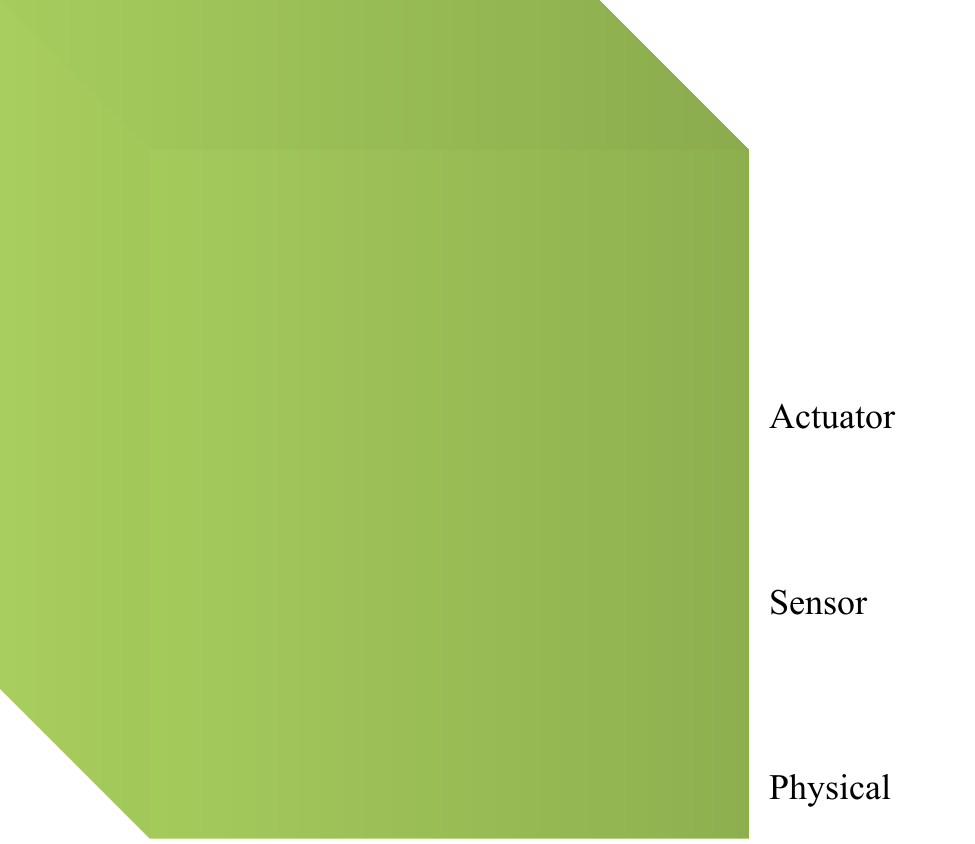}
\caption{Multilayered model of a \gls*{cps}.}
\label{fig:multi_layer_cps}
\end{figure}

The objective of our approach is twofold: (i)~A first step to achieving this goal is to ensure that the resilience potential of each layer of a given architecture is \textit{consistent}, i.e., ensuring that a layer is not resilient at the expense of the other ones. (ii)~The second step consists in protecting critical points. A critical point is an architecture's component, function, or subsystem. It is called \textit{critical} because an adversary attempting to attack a critical point can cause cascading effects that could generate important losses. According to Leveson \cite{leveson2011engineering}, a loss can be related to life or injury to people, damage to the material, the completion of the mission, the conformity of the regulations, reputation or finances. To achieve this goal, we consider the multilayered representation shown in Fig.~\ref{fig:multi_layer_cps}.

The first level is the physical layer. This layer includes the physical components not playing a role in a system's steerability or monitorability potential, e.g., a tank or a pipe. The second layer is the sensor one. The potential for monitorability is the first pillar of resilience. The third layer is the actuator one, referring to the steerability potential (the second pillar of resilience), including pumps and valves. Then, the cyber layer includes the components connected to cyberspace, i.e., sensors sending readings to a controller through a network. These connected components are visible to an adversary spying on them from cyberspace and attempting to inject data to put the system in an unstable state. It includes all the components sending data through a network. The mission layer corresponds to the components used to complete a mission of the system. We must note that the links connecting the nodes differ in the five identified layers. For example, a sensor link can be: \textit{Flowrate sensor sends data to the controller}. A mission link can be: \textit{Controller must check the water level in the tank according to the readings made by the level sensor}.

\begin{figure*}[ht]
\centering
    \begin{subfigure}[b]{0.6\textwidth}
        \includegraphics[width=\textwidth]{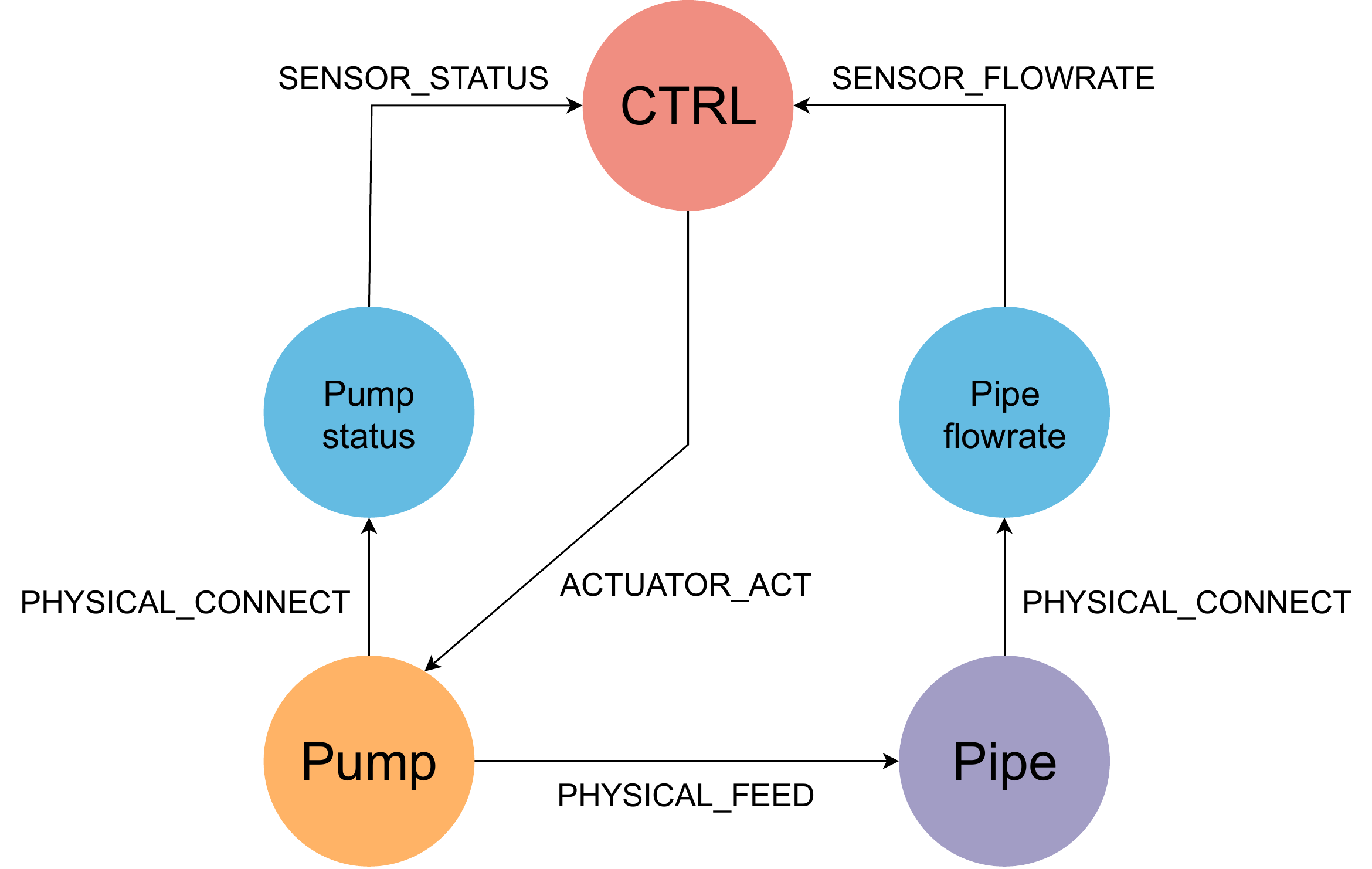}
        \caption{Semantic model in physical, sensor and actuator layers.}
        \medskip
        \label{fig:graph_semantic_psa}
    \end{subfigure}
    \begin{subfigure}[b]{0.45\textwidth}
        \includegraphics[width=\textwidth]{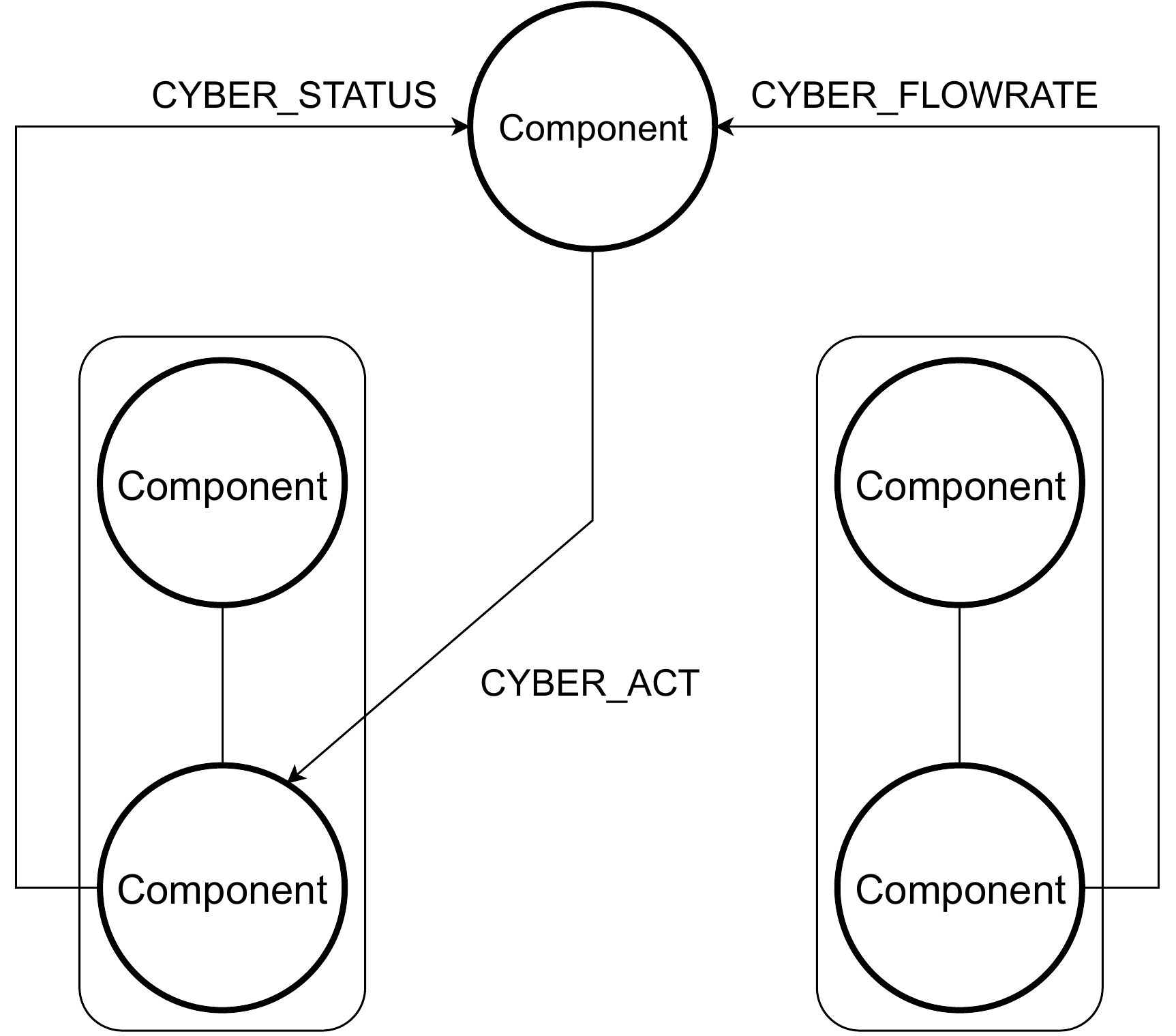}
        \caption{Semantic model in cyber layer.}
        \medskip
        \label{fig:graph_semantic_cyber}
    \end{subfigure}
    \begin{subfigure}[b]{0.54\textwidth}
        \includegraphics[width=\textwidth]{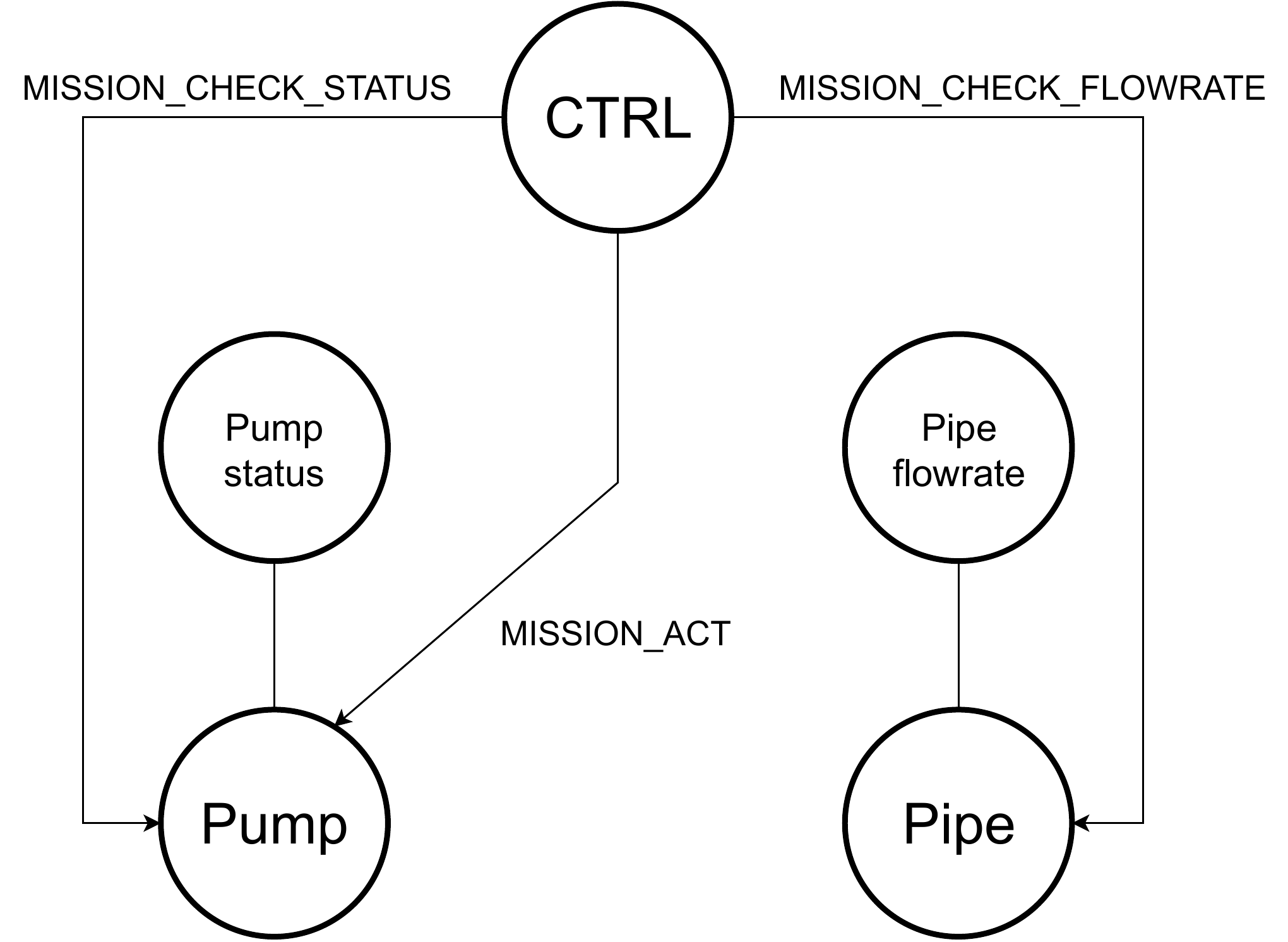}
        \caption{Semantic model in mission layer.}
        \medskip
        \label{fig:graph_semantic_mission}
    \end{subfigure}
    \caption{Semantic model inspired from a water treatment subsystem.}
    \label{fig:semantic_model}
\end{figure*}

Our layered model is based on mapping components according to the resilience potential they can bring to an architecture. The two last layers (cyber and mission) are transversal layers that cover the entire architecture. Fig.~\ref{fig:semantic_model} presents a semantic graph of a water treatment subsystem represented according to our multilayered model. 

Figs.~\ref{fig:semantic_model}(a), \ref{fig:semantic_model}(b) and \ref{fig:semantic_model}(c) represent the same graph used as an example with specific information related to the physical-sensor-actuator, cyber- and mission layers. The nodes are used to represent components of the system. The links between the nodes represent logical, physical, and communication exchanges. This model is suitable for resilience assessment and attack propagation analysis. The physical, sensor, and actuator layers map components according to their role in improving resilience. We remind the reader that monitorability and steerability are the two pillars of resilience, which are, respectively, related to the sensor and actuator layers. The cyber layer models the numerical dimension of components such as data exchanges between sensors and controllers, or control actions sent by a controller. The cyber layer corresponds to the knowledge that a cyber adversary interfering with the network can acquire by analyzing data exchange on a network. In Fig.~\ref{fig:semantic_model}(b), we modeled that the adversary cannot identify components. However, data analysis allows the adversary to build its model to estimate which components interact with each other. The mission layer models what is expected from each element to ensure the system can complete its mission with enough security measures to guarantee resilience.

\subsubsection{Resilience Quantification Using Graph Analytics Metrics}
\label{sec:metrics}

\noindent Modeling complex systems with knowledge graphs implies representing \gls*{it}/\gls*{ot} components by the bias of nodes. These nodes interact with each other through a set of links, representing physical, wireless, and logical relationships. These knowledge graph models allow us to find critical points, i.e., components or functions that can have a major impact on the performance of systems in case of a failure or when an attack occurs. An adversary attempting to target a critical point can significantly damage a system. Thus, identifying and protecting these critical points is paramount to ensure the resilience of \glspl*{cps}. Several metrics exist in graph analytics and can be relevant for resilience quantification purposes.

As mentioned by Newman, the Eigenvector Centrality metric measures the influence of neighbors on a node~\cite{newman2008mathematics,neo4j2024eigenvector}. 
Neighbors with high eigenvector centrality carry more weight in the measure than low-value neighbors.
A node with high eigenvector centrality is in relationships with several neighbors also having high eigenvector centrality.

In graph theory, a graph $G$ is defined by as $G=(V,E)$ where $V$ is a set of nodes and $E$ is a set of links between nodes \cite{bollobas1998modern}. A multilayered graph structure, also known in the literature as a multilayered $N$ dimensional network, is modeled with $M=(G,C)$ where $G=\{G_\alpha, \alpha \in \{1,...N\}\}$ is a set of graphs $G_\alpha=(V_\alpha,E_\alpha)$ and $C=\{E_{\alpha\beta}\subseteq X_\alpha \times X_\beta; \alpha,\beta \in \{1,...,N\}, \alpha \neq \beta\}$ is a set of links between $G_\alpha$ and $G_\beta$, where we have $\alpha \neq \beta$.

For a given graph $G=(V,E)$ with $|V|$ vertices, let $A=(a_{v,t})$ be the adjacency matrix, i.e., we have $a_{v,t}=1$ if the vertex $v$ is linked to the vertex $t$, and $a_{v,t}=0$ otherwise. The eigenvector centrality score of $v$ is:
\begin{equation}
    x_v = \frac{1}{\lambda} \sum_{t \in M(v)} x_t = \frac{1}{\lambda} \sum_{t \in G} a_{v,t}x_t
    \label{eq:eigenvector_centrality}
\end{equation}
with $M(v)$ the set of neighbors of $v$ and $\lambda$ a constant. Following the Newman reasoning \cite{newman2008mathematics}, Eq.~\eqref{eq:eigenvector_centrality} can be rewritten as follows $AX = \lambda X$. $X$ is an eigenvector of the adjacency matrix $A$, with eigenvalue $\lambda$. $\lambda$ must be the largest eigenvalue of the adjacency matrix $A$. According to the Perron-Frobenius theorem, this choice guarantees that if $A$ is irreducible, i.e., if the considered graph is (strongly) connected, then the eigenvector solution $X$ is unique and positive. Such a metric is interesting for catching the influence of neighbor nodes, which is related to the notion of critical point. A critical point or node in a graph model is an important function, component, or subsystem for completing a mission. 
However, a critical point could also generate cascading effects when an adversary attempts to attack it. This notion of a critical point is important in multilayered models. A layer's general resilience is insufficient to ensure a good resilience potential. We must ensure that an adversary cannot target critical points to generate cascading effects.

Another interesting metric for resilience purposes is the Betweenness Centrality. It is used to detect a node's influence on the information flow that navigates through a graph \cite{neo4j2024betweenness}. A high betweenness centrality value implies that a node serves as a bridge from one part of a graph to another in which a large amount of information is exchanged. Brandes~\etal specified that the algorithm works for positively weighted (with multiple concurrent Dijkstra algorithms) and nonweighted graphs (with the Brandes' approximate algorithm \cite{brandes2007centrality}). The implementation requires $O(n + m)$ space and runs in $O(n \cdot m)$ time for a graph $G$, with $n$ the number of nodes and $m$ the number of relationships in $G$. This metric can capture the importance of nodes according to the amount of information received. Thus, identifying critical points could be possible.

A third interesting metric is the \gls*{wcc} algorithm, which finds sets of connected nodes in directed and undirected graphs. Two nodes are connected if a path exists between them. A set of connected nodes is said to be a component \cite{neo4j2025wcc}. These communities, or components, can be viewed as groups of critical points. This metric can also provide interesting results in the identification of critical points.

\subsection{Related Work}
\label{sec:related_work}

\noindent In the sequel, we survey some work related to graph techniques used in cybersecurity.

\subsubsection{Graph Techniques in Cybersecurity}
\paragraph{Graph Analytics}

Graph analytics is a data analysis used to understand complex relationships between data entities represented in a graph. It consists of evaluating information and their connections to determine how the elements are or could be related. Noel reviews graph-based methods to assess and improve operational computer network security, maintain situational awareness, and ensure organizational missions \cite{noel2018review}.

\paragraph{Graph Mining}

Securing cyberspace and exchanging sensitive data have become paramount for organizations, governments, and industrial firms. Graph mining is a set of techniques used for different purposes: (i)~conduct analysis about the properties of real-world graphs; 
(ii)~understand and establish predictions about how a graph can affect some application, and (iii)~build models to generate realistic graphs that match real world graph patterns. Building on graph mining techniques created by the scientific community, researchers are trying to capture correlations between cyber entities. The work by Yan~\etal \cite{yan2023graph} presents a review of graph mining techniques used for cybersecurity.

\paragraph{Graph Visualization}

Analyzing complex graph structures requires visualization tools with integrated plugins for graph analytics. Several tools for graph visualization analysis are available, such as Gephi. Bastian~\etal describe it as a graph visualization software initially dedicated to social network analysis \cite{bastian2009gephi}. The Arena 3D Web tool is also interesting because it considers multidimensional structures. Kokoli~\etal present it as an interactive application that allows us to visualize graphs modeled as multilayered structures in 3D space \cite{kokoli2023arena3dweb}. In our resilience assessment, we use the Neo4j software, for which Scifo presents a description \cite{scifo2023graph}. We use it for its packages, including graph analytics algorithms and metrics. This tool is also relevant for our analysis due to its \gls*{llm} tool that we used in Section~\ref{sec:experimentation}.

\subsubsection{Multilayered Approaches}

\noindent Modeling systems as multilayered architectures is not a new topic in the literature. Gardner introduced two multilevel approaches for modeling systems with the SARA design, considering \textit{relatively abstract submodels} \cite{gardner1977multi}. Before this work, Zurcher~\etal highlighted the importance of considering the levels of abstraction in modeling strategies \cite{zurcher1968iterative}. Zurcher's work introduces a technique for modeling a multiprocessing system's hardware and software components. More recently, Carreras~\etal presented an approach to consider the key features of \glspl*{cps} by the bias of a multilayered representation for safety and security analysis purposes \cite{carreras2020conceptualizing}.

Multilayered representations are also pyramidal representations to model a system's architectural, logical, or regulation-related levels.

There are frameworks, i.e., the \gls*{iira} presented by Lin~\etal \cite{lin2015iira} and \gls*{rami} from Hankel~\etal work \cite{hankel2015rami}, suitable for modeling Industry 4.0 architectures as multilayered systems. \gls*{rami} uses a 3-D model by representing an architecture with the following layers: asset, integration, communication, information, functional, and business. In our previous work \cite{dagnas2024methodological}, we applied the \gls*{rami} model to a water treatment architecture.

\subsubsection{Attack Graphs for Resilience Purposes}

\noindent In their work, Al Ghazo and Kumar proposed a methodology to identify critical attacks that could compromise the behavior of a system and, when blocked, to guarantee the security of the system \cite{alghazo2019identification}. Zonouz~\etal work on a different axis. Indeed, their work is based on contingency analysis, which provides guidelines to achieve resilience goals and allows a system to continue to operate even if a failure occurs. In addition to this methodology, they propose using a cyber-physical security evaluation technique that plans remediation measures for accidental and intentional adversarial events \cite{zonouz2014security}. This methodology can help operators choose prevention solutions in case of proactive intrusions. However, such a technique works before an adversarial event occurs and does not increase a system's resilience potential because a human operator's action is required. 
Furthermore, the system cannot return to a stable state when an adversary can bypass these measures.

We learn from works in the literature that resilience strategies are rarely involved and associated to graph analytics for quantification and enhancement purposes. The modeling of complex systems remains a significant problem. Most metrics used to quantify resilience are closely related to the models used and cannot be transposed to other models. Knowledge graphs are powerful for modeling complex systems because of their ability to consider various diversities of entities and connections between them.

To address the stated challenges, we propose to associate our multilayered model introduced in our previous work \cite{dagnas2024multi} based on knowledge graphs to graph analytics metrics to provide a resilience assessment of three \gls*{swat} designs. Our methodology highlights critical points that need to be protected from cyber adversaries.

\section{A Resilience Assessment of SWaT}
\label{sec:swatra}

\noindent We consider the \gls*{swat} testbed as a use case to perform our resilience assessment. \gls*{swat} was released by the \gls*{sutd}. This system mimics the real behavior of the Singapore water treatment facility. The water treatment case is relevant to our analysis. Firstly, it illustrates the critical aspect of a mission. In the event of an attack on a water treatment facility, the health of the people who drink the water is directly affected. Secondly, this example aligns with current events, like the cyberattack perpetrated against the Florida water treatment plant in 2021 and reported by Greenberg in \cite{greenberg2021hacker}.

\gls*{swat} is described in the work by Goh~\etal \cite{goh2017dataset}. The system is divided into six stages: \textit{Pumping}, \textit{Chemical Dosing}, \textit{\gls*{uf}}, \textit{Dechlorination}, \textit{\gls*{ro}}, \textit{Final stage and Backwash of the \gls*{uf} membrane}. As a use case, we consider the first stage of \gls*{swat}, in which raw water must be cleaned and pumped into the system.

\begin{figure*}[!htp]
\centering
    \begin{subfigure}[b]{0.9\textwidth}
        \includegraphics[width=\textwidth]{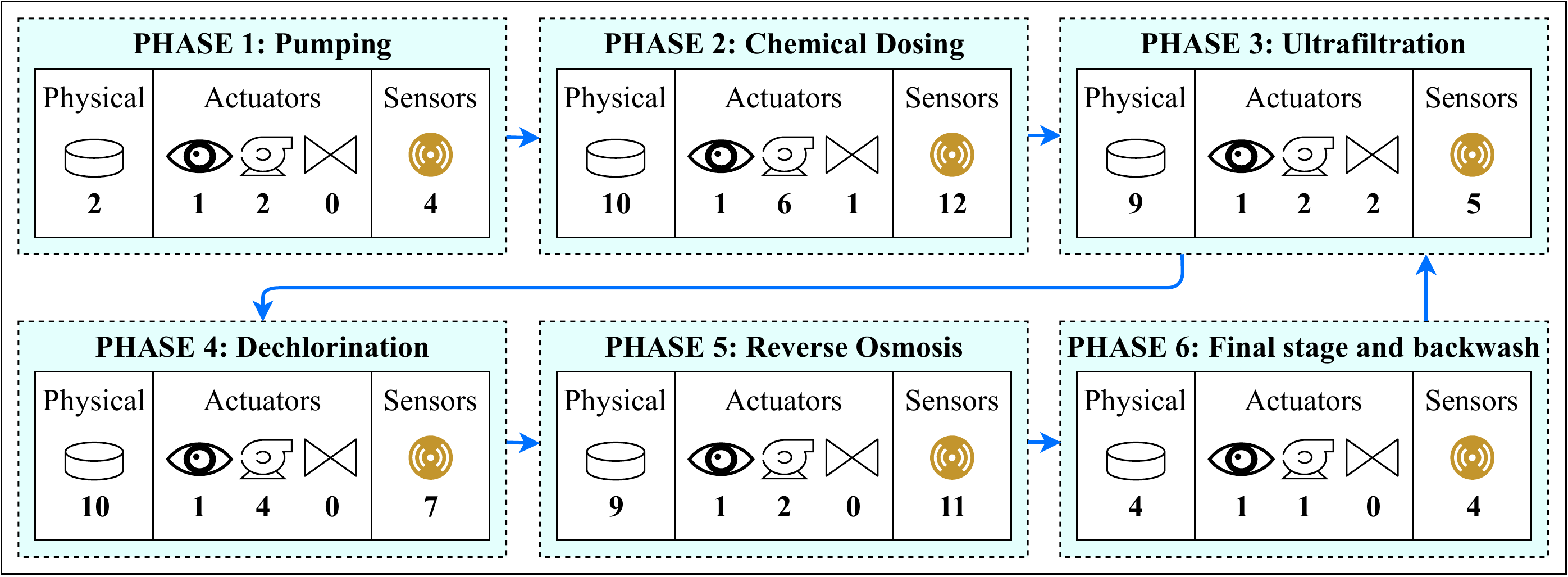}
        \caption{\gls*{swat} original architecture.}
        \medskip
        \label{fig:swat_1_original}
    \end{subfigure}
    \begin{subfigure}[b]{0.9\textwidth}
        \includegraphics[width=\textwidth]{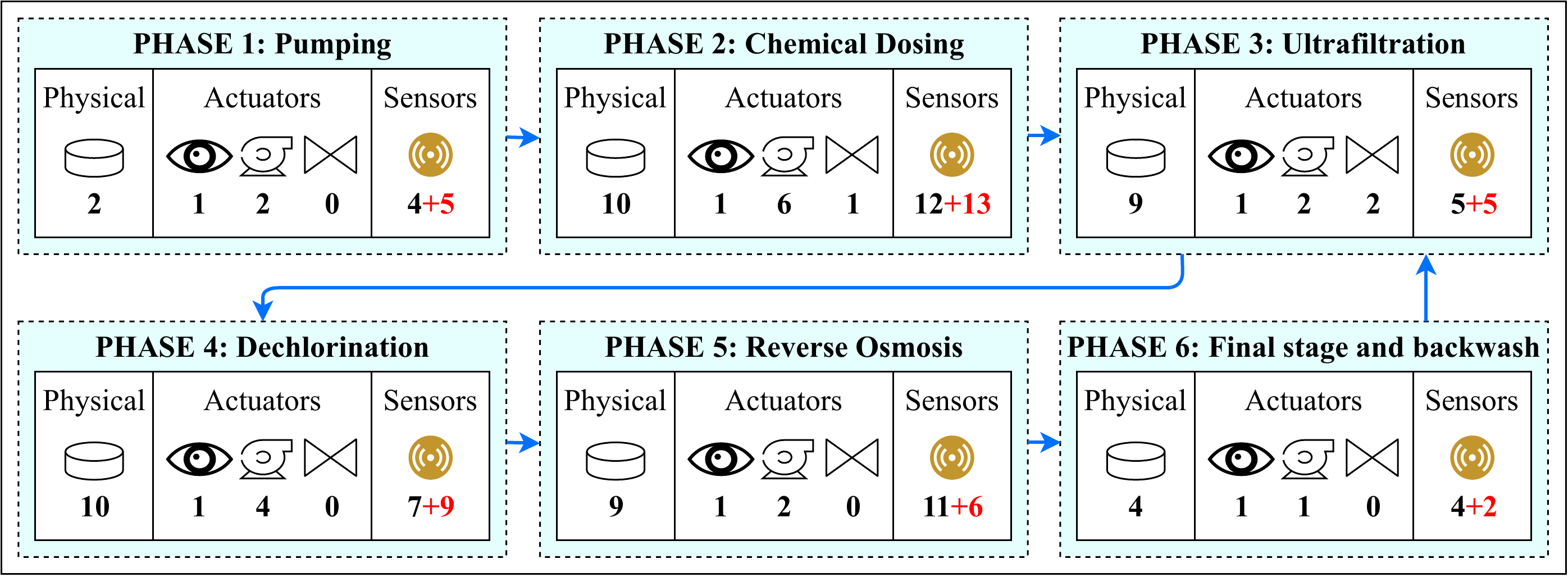}
        \caption{\gls*{swat} architecture with additional sensors.}
        \medskip
        \label{fig:swat_2_add_sensors}
    \end{subfigure}
    \begin{subfigure}[b]{0.9\textwidth}
        \includegraphics[width=\textwidth]{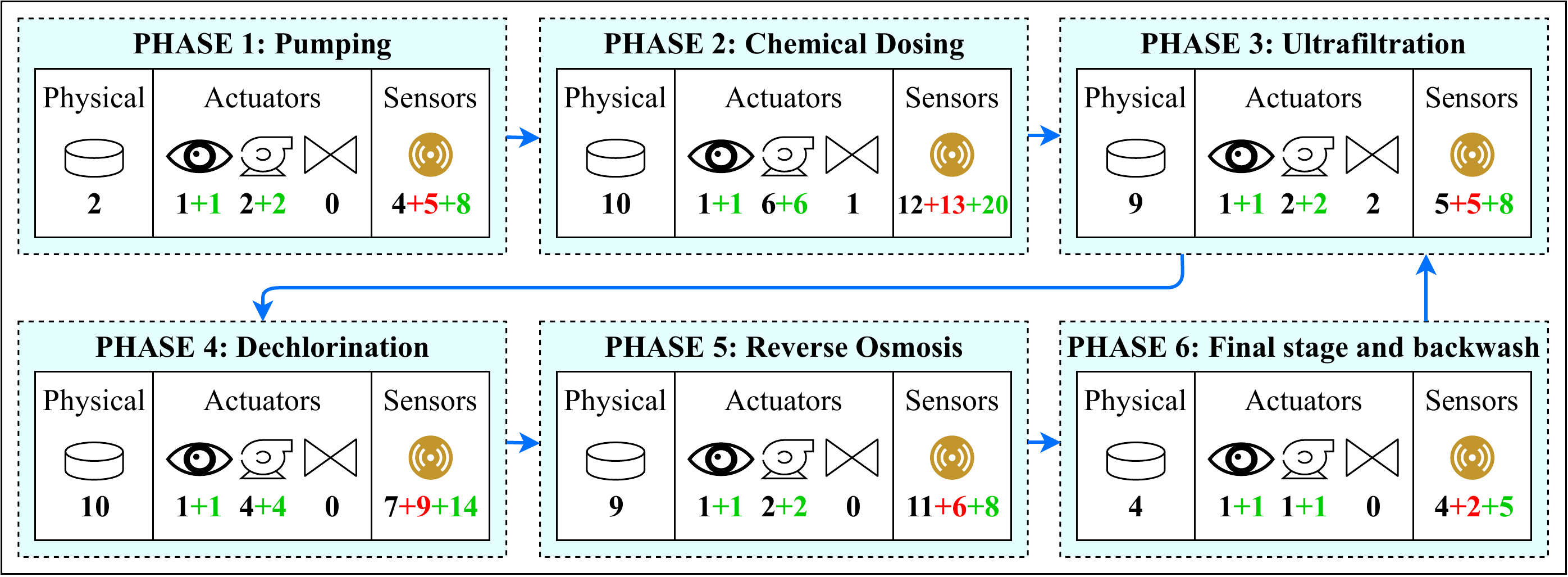}
        \caption{\gls*{swat} architecture with additional sensors and additional actuators.}
        \medskip
        \label{fig:swat_3_add_sensors_and_act}
    \end{subfigure}
    \begin{subfigure}[b]{0.9\textwidth}
        \includegraphics[width=\textwidth]{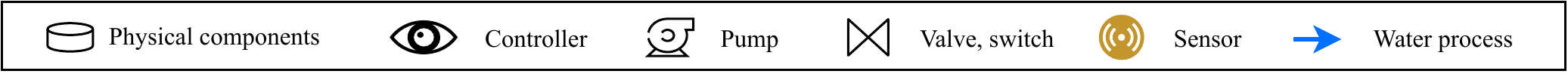}
    \end{subfigure}
    \caption{Alternative designs of \gls*{swat} for resilience evaluation purposes. Fig.~\ref{fig:swat_1_original} represents $A_1$, the original \gls*{swat} design. In Fig.~\ref{fig:swat_2_add_sensors}, $A_2$ is similar to $A_1$, with additional sensors that increase the monitorability potential. Fig.~\ref{fig:swat_3_add_sensors_and_act} representing $A_3$ includes redundant controllers with auxiliary actuators, which increases the steerability capacities of the system. Monitorability has also increased compared to $A_2$ because additional actuators also need to be monitorable. }
\end{figure*}

Fig.~\ref{fig:swat_1_original} presents the original \gls*{swat} architecture, including the numbers of components related to the physical, sensor, and actuator layers. Fig~\ref{fig:swat_2_add_sensors} presents the same architecture with additional sensors, represented in red. The monitorability potential has increased. In the case of Fig.~\ref{fig:swat_3_add_sensors_and_act}, we consider the architecture presented in Fig.~\ref{fig:swat_2_add_sensors} for which the steerability potential has been improved with additional actuators and controllers. Thus, because of the increase in steerability potential, a new group of sensors must also be included to extend the monitorability capacities of the architecture.

Building these figures follows the resilience postulate. Indeed, a gain in the resilience potential of a system implies an increase in the monitorability degree, i.e., by adding sensors, and the steerability degree, i.e., by adding actuators. By applying the previously introduced metrics, namely eigenvector centrality, betweenness centrality and weakly connected components, we quantify the resilience of the three multilayered architectures (Figs.~\ref{fig:swat_1_original}, \ref{fig:swat_2_add_sensors}, and \ref{fig:swat_3_add_sensors_and_act}). The objective of this analysis is twofold: (i) We quantify the resilience of these three architectures with the three presented metrics; (ii) We compare the results obtained with each metric and discuss the accuracy of the obtained results.

The results obtained with our multilayered assessment are available in the spreadsheet placed in our Github repository\footnote{Github repository available at: \url{https://github.com/romaindgns/cyber_resilience_analytics}} \cite{github_cyberresilienceanalytics}. The observation made is that the Eigenvector centrality gives results with a fine granularity in comparison with the Betweenness centrality or with \gls*{wcc}. According to our results, the Betweenness centrality identifies most of the critical points highlighted by the Eigenvector centrality. However, there is a loss of information for the sensor layer of our model in which all the components obtain a zero score. In addition, the cyber and mission layers obtained very similar results with the Betweenness centrality assessment. However, the Eigenvector centrality makes a clear distinction between these two layers. Consider the example of the \gls*{swat} original design pumping stage \gls*{swat}. Critical nodes in the cyber layer are the controllers, which is in accordance with their high influence in receiving readings and sending control actions. In the mission layer, the identified critical points are the pumps in charge of driving water through the system. This distinction is not appearing with the Betweenness centrality. The \gls*{wcc} allows identifying components, i.e., sets of nodes in a graph. This metric identifies groups between physical components and sensors in our designs. However, the obtained results are not sufficiently accurate to identify critical points. This is why we recommend considering the Eigenvector centrality for critical points identification.

\begin{figure}[!htp]
\centering
\begin{subfigure}[b]{0.49\textwidth}
    \includegraphics[width=\textwidth]{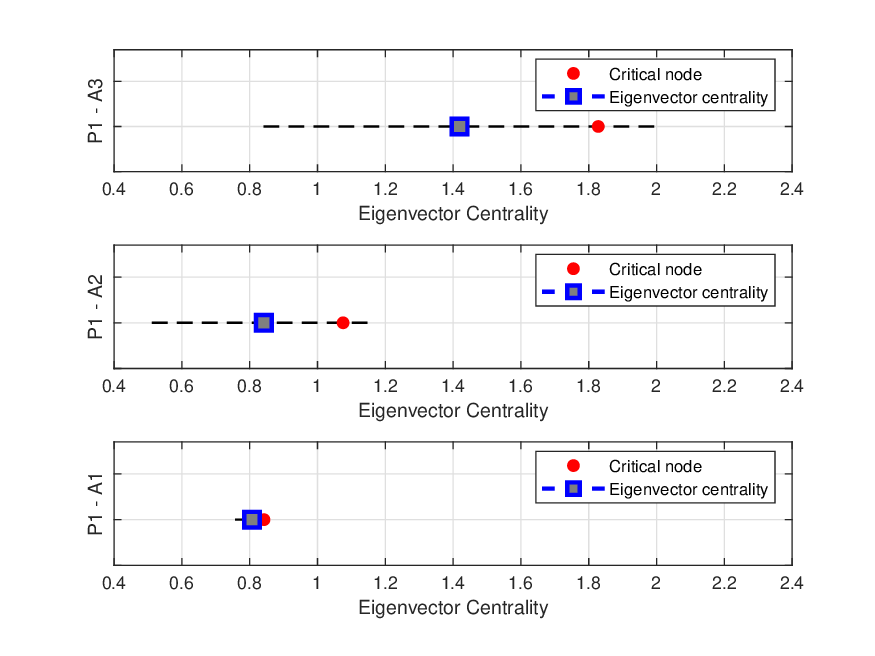}
    \caption{\gls*{swat} pumping stage - Physical layer.}
    \medskip
    \label{fig:swat_p1_physical}
\end{subfigure}
\begin{subfigure}[b]{0.49\textwidth}
    \includegraphics[width=\textwidth]{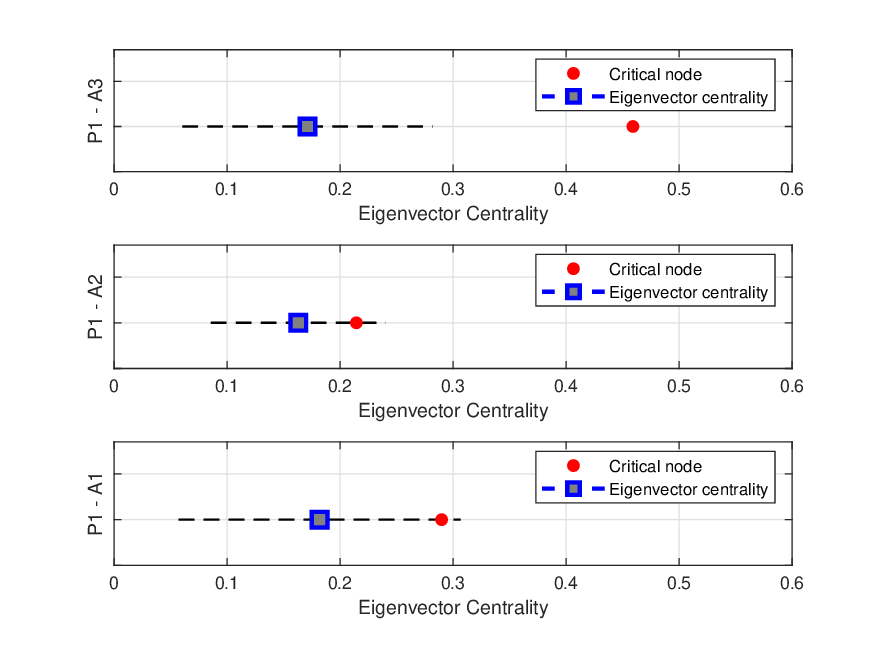}
    \caption{\gls*{swat} pumping stage - Sensor layer.}
    \medskip
    \label{fig:swat_p1_sensor}
\end{subfigure}
\begin{subfigure}[b]{0.49\textwidth}
    \includegraphics[width=\textwidth]{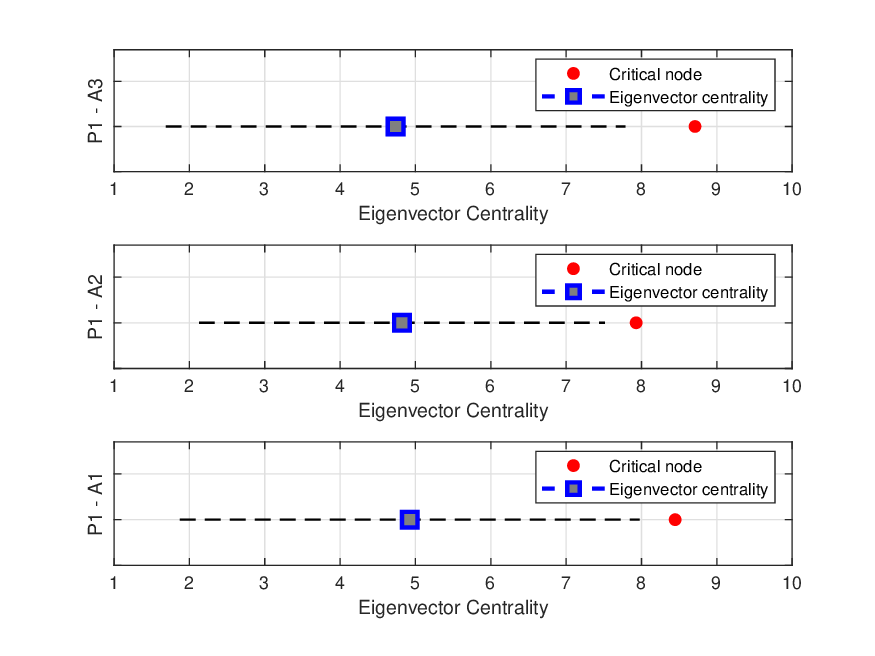}
    \caption{\gls*{swat} pumping stage - Actuator layer.}
    \medskip
    \label{fig:swat_p1_actuator}
\end{subfigure}
\begin{subfigure}[b]{0.49\textwidth}
    \includegraphics[width=\textwidth]{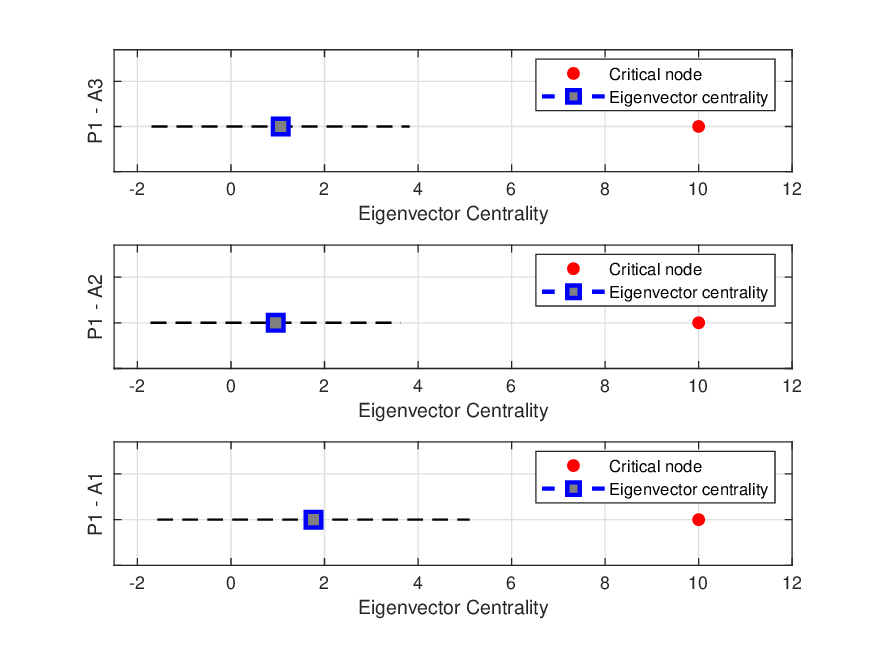}
    \caption{\gls*{swat} pumping stage - Cyber layer.}
    \medskip
    \label{fig:swat_p1_cyber}
\end{subfigure}
\begin{subfigure}[b]{0.49\textwidth}
    \includegraphics[width=\textwidth]{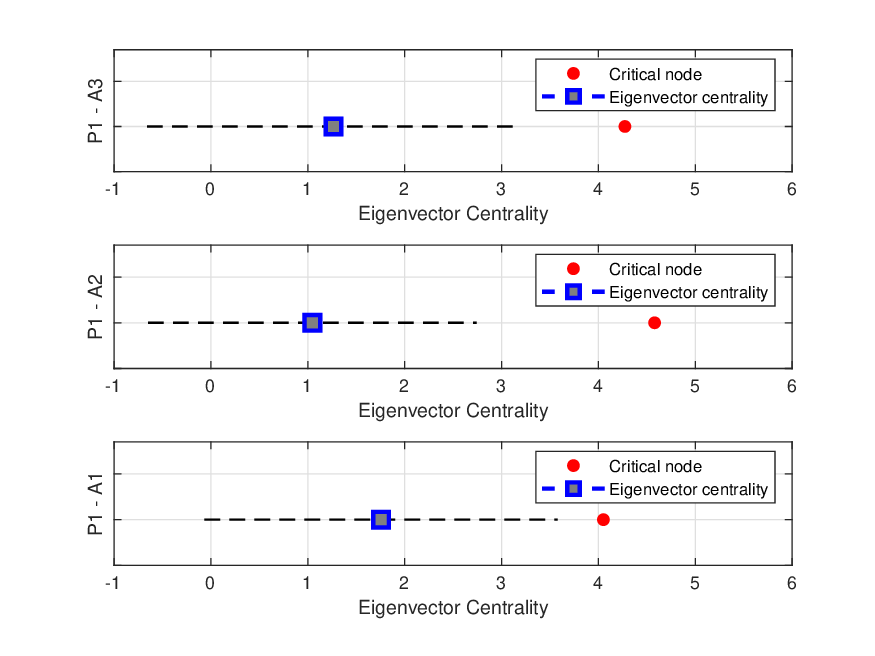}
    \caption{\gls*{swat} pumping stage - Mission layer.}
    \medskip
    \label{fig:swat_p1_mission}
\end{subfigure}

\caption{Resilience assessment of \gls*{swat} pumping stage designs.}
\label{fig:swat_p1_eig}
\end{figure}

Figures presenting a graphical representation of the eigenvector centrality results for the three designs of \gls*{swat} pumping stage are presented in Figure~\ref{fig:swat_p1_eig}. Figures presenting the results for the five other \gls*{swat} subsystems are available in our GitHub repository \cite{github_cyberresilienceanalytics}. Mean eigenvector centrality results computed with the values obtained for each component in each layer are depicted in blue squares. Dark dashed lines are used to represent the \gls*{std}. Red points are the critical ones, identified in each layer by having the maximum eigenvector centrality value. 

\begin{table}[!ht]
\footnotesize
\setlength{\tabcolsep}{4pt}
\renewcommand{\arraystretch}{0.95}
\begin{center}
\caption{\gls*{swat} original design - Critical points}

\begin{tabular}{ c|c|c|c|c|c| }
\cline{2-6}
&
\textbf{Physical} &
\textbf{Sensor} &
\textbf{Actuator} &
\textbf{Cyber} &
\textbf{Mission} \\
\hline
\multicolumn{1}{ |l| }{\multirow{1}{*}{Stage 1.}} & 
  \multirow{2}{*}{Pipe1}&
  Pump1\_status&
  \multirow{2}{*}{CTRL1}&
  \multirow{2}{*}{CTRL1}&
  Pump1\\
 \multicolumn{1}{ |l| }{\multirow{1}{*}{Pumping}} &
  &
  Pump2\_status&
    &
  &
 Pump2\\
 \cline{1-6}
\multicolumn{1}{ |l| }{\multirow{2}{*}{Stage 2.}} &
  \multirow{4}{*}{Mixer} &
  Pump$_x$\_NaCl\_status &
  \multirow{4}{*}{CTRL2} &
  \multirow{4}{*}{CTRL2} &
  Pump$_x$\_NaCl\\
 \multicolumn{1}{ |l| }{\multirow{2}{*}{Chemical}} &
 &
 Pump$_x$\_NaOCl\_status &
   &
  &
  Pump$_x$\_NaOCl\\
 \multicolumn{1}{ |l| }{\multirow{2}{*}{Dosing}} &
 &
 Pump$_x$\_HCl\_status &
   &
  &
  Pump$_x$\_HCl\\
 \multicolumn{1}{ |l| }{} &
 &
 \tiny{with $x \in \{1,2\}$} &
   &
  &
  \tiny{with $x \in \{1,2\}$}\\
\cline{1-6}
\multicolumn{1}{ |l| }{\multirow{1}{*}{Stage 3.}} &
  \multirow{2}{*}{Membrane}&
  Pump1\_status&
  \multirow{2}{*}{CTRL3} &
  \multirow{2}{*}{CTRL3} &
  Pump1\\
 \multicolumn{1}{ |l| }{\multirow{1}{*}{\gls*{uf}}} &
  &
  Pump2\_status&
   &
  &
  Pump2\\
 \cline{1-6}
\multicolumn{1}{ |l| }{\multirow{1}{*}{Stage 4.}} &
  \multirow{3}{*}{UV Unit}&
  \multirow{2}{*}{Pump1\_status}&
  \multirow{3}{*}{CTRL4} &
  \multirow{3}{*}{CTRL4} &
  Pump$_x$\\
 \multicolumn{1}{ |l| }{\multirow{1}{*}{Dechlori-}} &
 &
  \multirow{2}{*}{Pump2\_status}&
   &
  &
 Pump$_x$\_$\text{NaHSO}_3$\\
\multicolumn{1}{ |l| }{\multirow{1}{*}{nation}} &
 &
 &
   &
  &
  \tiny{with $x \in \{1,2\}$}\\
   \cline{1-6}
\multicolumn{1}{ |l| }{\multirow{2}{*}{Stage 5.}} &
  \multirow{3}{*}{\gls*{ro} Unit}&
  \multirow{2}{*}{PumpBoost1\_status}&
  \multirow{3}{*}{CTRL5} &
  \multirow{3}{*}{CTRL5} &
  Pipe6\\
 \multicolumn{1}{ |l| }{\multirow{2}{*}{\gls*{ro}}} &
 &
  \multirow{2}{*}{PumpBoost2\_status}&
   &
  &
  PumpBoost1\\
  \multicolumn{1}{ |l| }{} &
 &
  &
   &
  &
  PumpBoost2\\
 \cline{1-6}
\multicolumn{1}{ |l| }{\multirow{1}{*}{Stage 6.}} &
  \multirow{2}{*}{Pipe3}&
  \multirow{2}{*}{PumpBack\_status}&
  \multirow{2}{*}{CTRL6}&
  \multirow{2}{*}{CTRL6}&
  \multirow{2}{*}{Pump\_Backwash}\\
 \multicolumn{1}{ |l| }{\multirow{1}{*}{Backwash}} &
  &
  &
    &
  &
 \\
\hline

\end{tabular}
\label{tab_critical_point_a1}
\end{center}
\end{table}

\begin{table}[!ht]
\footnotesize
\setlength{\tabcolsep}{4pt}
\renewcommand{\arraystretch}{0.95}
\begin{center}
\caption{\gls*{swat} original design with additional sensors - Critical points}

\begin{tabular}{ c|c|c|c|c|c| }
\cline{2-6}
&
\textbf{Physical} &
\textbf{Sensor} &
\textbf{Actuator} &
\textbf{Cyber} &
\textbf{Mission} \\
\hline
\multicolumn{1}{ |l| }{\multirow{2}{*}{Stage 1.}} & 
  \multirow{3}{*}{Tank1}&
  Pump$_{x,y}$ &
  \multirow{3}{*}{CTRL1}&
  \multirow{3}{*}{CTRL1}&
  \multirow{2}{*}{Pump1}\\
 \multicolumn{1}{ |l| }{\multirow{2}{*}{Pumping}} &
  &
  \tiny{with $x \in \{1,2\}$} &
    &
  &
 \multirow{2}{*}{Pump2}\\
   \multicolumn{1}{ |l| }{} &
  &
  \tiny{$y \in $\{status, temp, rotation\}}&
  &
  &
 \\
   \hline
\multicolumn{1}{ |l| }{\multirow{3}{*}{Stage 2.}} &
  \multirow{5}{*}{Mixer} &
  \multirow{1}{*}{Pump$_x$\_NaCl$_y$} &
  \multirow{5}{*}{CTRL2} &
  \multirow{5}{*}{CTRL2} &
  \multirow{2}{*}{Pump$_x$\_NaCl} \\
 \multicolumn{1}{ |l| }{\multirow{3}{*}{Chemical}} &
 &
  \multirow{1}{*}{Pump$_x$\_NaOCl$_y$} &
   &
  &
   \multirow{2}{*}{Pump$_x$\_NaOCl} \\
 \multicolumn{1}{ |l| }{\multirow{3}{*}{Dosing}} &
 &
  \multirow{1}{*}{Pump$_x$\_HCl$_y$} &
   &
  &
   \multirow{2}{*}{Pump$_x$\_HCl} \\
 \multicolumn{1}{ |l| }{} &
 &
 \multirow{1}{*}{\tiny{with $x \in \{1,2\}$}} &
   &
  &
  \multirow{2}{*}{\tiny{with $x \in \{1,2\}$}} \\
 \multicolumn{1}{ |l| }{} &
 &
 \multirow{1}{*}{\tiny{$y \in \{status, temp, rotation$\}}} & 
   &
  &
  \\
 \cline{6-6}
\cline{1-6}
\multicolumn{1}{ |l| }{\multirow{2}{*}{Stage 3.}} &
  \multirow{3}{*}{Membrane}&
  \multirow{1}{*}{Pump$_{x,y}$} &
  \multirow{3}{*}{CTRL3} &
  \multirow{3}{*}{CTRL3} &
  \multirow{2}{*}{Pump1}\\
 \multicolumn{1}{ |l| }{\multirow{2}{*}{\gls*{uf}}} &
  &
  \multirow{1}{*}{\tiny{with $x \in \{1,2\}$}} &
   &
  &
  \multirow{2}{*}{Pump2}\\
 \multicolumn{1}{ |l| }{} &
  &
  \multirow{1}{*}{\tiny{$y \in \{status, temp, rotation$\}}} & 
   &
  &
  \\
 \cline{1-6}
\multicolumn{1}{ |l| }{\multirow{1}{*}{Stage 4.}} &
  \multirow{3}{*}{UV Unit}&
  \multirow{1}{*}{Pump$_{x,y}$} &
  \multirow{3}{*}{CTRL4} &
  \multirow{3}{*}{CTRL4} &
  \multirow{1}{*}{Pump$_{x}$}\\
 \multicolumn{1}{ |l| }{\multirow{1}{*}{Dechlori-}} &
 &
  \multirow{1}{*}{\tiny{with $x \in \{1,2\}$}} &
   &
  &
 \multirow{1}{*}{Pump$_{x}$\_$\text{NaHSO}_3$}\\
\multicolumn{1}{ |l| }{\multirow{1}{*}{nation}} &
 &
 \multirow{1}{*}{\tiny{$y \in \{status, temp, rotation$\}}} & 
   &
  &
  \multirow{1}{*}{\tiny{with $x \in \{1,2\}$}}\\
   \cline{1-6}
\multicolumn{1}{ |l| }{\multirow{1}{*}{Stage 5.}} &
  \multirow{3}{*}{\gls*{ro} Unit}&
  Pumpboost$_{x,y}$ &
  \multirow{3}{*}{CTRL5} &
  \multirow{3}{*}{CTRL5} &
  \multirow{2}{*}{PumpBoost1}\\
 \multicolumn{1}{ |l| }{\gls*{ro}} &
  &
  \tiny{with $x \in \{1,2\}$} &
   &
  &
  \multirow{2}{*}{PumpBoost2}\\
\multicolumn{1}{ |l| }{} &
  &
  \tiny{$y \in $\{status, temp, rotation\}}&
   &
  &
  \\
 \cline{1-6}
\multicolumn{1}{ |l| }{\multirow{1}{*}{Stage 6.}} &
  \multirow{2}{*}{Pipe3}&
  \multirow{1}{*}{Pump\_backwash$_x$}&
  \multirow{2}{*}{CTRL6}&
  \multirow{2}{*}{CTRL6}&
  \multirow{2}{*}{Pump\_Backwash}\\
 \multicolumn{1}{ |l| }{\multirow{1}{*}{Backwash}} &
  &
  \multirow{1}{*}{\tiny{with $x \in $\{status, temp, rotation\}}} &
    &
  &
 \\
\hline

\end{tabular}
\label{tab_critical_point_a2}

\end{center}
\end{table}

\begin{table}[!ht]
\footnotesize
\setlength{\tabcolsep}{4pt}
\renewcommand{\arraystretch}{0.95}
\begin{center}
\caption{\gls*{swat} design with additional sensors and actuators - Critical points}

\begin{tabular}{ c|c|c|c|c|c| }
\cline{2-6}
&
\textbf{Physical} &
\textbf{Sensor} &
\textbf{Actuator} &
\textbf{Cyber} &
\textbf{Mission} \\
\hline
\multicolumn{1}{ |l| }{\multirow{2}{*}{Stage 1.}} & 
  \multirow{3}{*}{Tank1}&
  \multirow{3}{*}{CTRL1\_feedback}&
  \multirow{3}{*}{CTRL1}&
  \multirow{2}{*}{CTRL1}&
  \multirow{1}{*}{Tank1}\\
 \multicolumn{1}{ |l| }{\multirow{2}{*}{Pumping}} &
  &
  &
    &
   \multirow{2}{*}{CTRL1\_2} &
   \multirow{1}{*}{Pump$_x$}\\
\multicolumn{1}{ |l| }{} &
  &
  &
    &
    &
   \multirow{1}{*}{\tiny{with $x \in \{1,4\}$}}\\
 \cline{1-6}
\multicolumn{1}{ |l| }{\multirow{2}{*}{Stage 2.}} &
  \multirow{4}{*}{Mixer} &
  \multirow{4}{*}{CTRL2\_feedback} &
  \multirow{4}{*}{CTRL2} &
  \multirow{3}{*}{CTRL2} &
  \multirow{1}{*}{Mixer} \\
 \multicolumn{1}{ |l| }{\multirow{2}{*}{Chemical}} &
 &
  &
  &
  \multirow{3}{*}{CTRL2\_2} &
   \multirow{1}{*}{Pump$_{x,y}$} \\
\multicolumn{1}{ |l| }{\multirow{2}{*}{Dosing}} &
 &
  &
  &
   &
  \multirow{1}{*}{\tiny{with $x \in \{1,4\}$}} \\
     \multicolumn{1}{ |l| }{} &
 &
  &
  &
   &
  \multirow{1}{*}{\tiny{$y \in$ \{NaCl, NaOCl, HCl\}}} \\
\cline{1-6}
\multicolumn{1}{ |l| }{\multirow{2}{*}{Stage 3.}} &
  \multirow{3}{*}{Membrane}&
  \multirow{3}{*}{CTRL3\_feedback} &
  \multirow{3}{*}{CTRL3} &
  \multirow{2}{*}{CTRL3} &
  \multirow{1}{*}{UF\_membrane}\\
 \multicolumn{1}{ |l| }{\multirow{2}{*}{\gls*{uf}}} &
  &
  &
   &
  \multirow{2}{*}{CTRL3\_2} &
  \multirow{1}{*}{Pump$_x$}\\
   \multicolumn{1}{ |l| }{} &
  &
  &
   &
   &
  \multirow{1}{*}{\tiny{with $x \in \{1,4\}$}}\\
 \cline{1-6}
\multicolumn{1}{ |l| }{\multirow{2}{*}{Stage 4.}} &
  \multirow{4}{*}{UV Unit}&
  \multirow{4}{*}{CTRL4\_2\_feedback}&
  \multirow{4}{*}{CTRL4\_2} &
  \multirow{3}{*}{CTRL4} &
  \multirow{1}{*}{UV Unit}\\
 \multicolumn{1}{ |l| }{\multirow{2}{*}{Dechlori-}} &
 &
  &
   &
  \multirow{3}{*}{CTRL4\_2}&
 \multirow{1}{*}{Pump$_x$} \\
  \multicolumn{1}{ |l| }{\multirow{2}{*}{nation}} &
 &
  &
   &
  &
 \multirow{1}{*}{Pump$_x$\_$\text{NaHSO}_3$} \\
\multicolumn{1}{ |l| }{} &
 &
  &
   &
  &
 \multirow{1}{*}{\tiny{with $x \in \{1,4\}$}} \\
 \cline{1-6}

\multicolumn{1}{ |l| }{\multirow{3}{*}{Stage 5.}} &
  \multirow{4}{*}{\gls*{ro} Unit}&
  \multirow{4}{*}{CTRL5\_2\_feedback}&
  \multirow{4}{*}{CTRL5\_2} &
  \multirow{3}{*}{CTRL5} &
  \multirow{1}{*}{\gls*{ro} Unit}\\
 \multicolumn{1}{ |l| }{\multirow{3}{*}{\gls*{ro}}} &
  &
  &
   &
  \multirow{3}{*}{CTRL5\_2}&
  \multirow{1}{*}{Pipe6}\\
   \multicolumn{1}{ |l| }{} &
  &
  &
   &
  &
  \multirow{1}{*}{Pumpboost$_x$} \\
   \multicolumn{1}{ |l| }{} &
  &
  &
   &
  &
  \multirow{1}{*}{\tiny{with $x \in \{1,4\}$}}\\
 \cline{1-6}
\multicolumn{1}{ |l| }{\multirow{1}{*}{Stage 6.}} &
  \multirow{2}{*}{Pipe3}&
  \multirow{2}{*}{CTRL6\_2\_feedback}&
  \multirow{2}{*}{CTRL6\_2}&
  \multirow{1}{*}{CTRL6}&
  \multirow{1}{*}{Pump\_Backwash}\\
 \multicolumn{1}{ |l| }{\multirow{1}{*}{Backwash}} &
  &
  &
    &
  \multirow{1}{*}{CTRL6\_2}&
  \multirow{1}{*}{Pump\_Backwash\_aux}\\
\hline

\end{tabular}
\label{tab_critical_point_a3}
\end{center}
\end{table}

The identified critical points for the three \gls*{swat} designs, according to the Eigenvector centrality results presented in Tables~\ref{tab_critical_point_a1}, \ref{tab_critical_point_a2}, and \ref{tab_critical_point_a3}. According to the Eigenvector centrality results, the critical points stay the same in each layer across the designs. The metric clearly identifies the components that play an important role as the most critical ones. It is interesting to see that in the cyber layer, controllers are considered critical points, and in the mission layer, pumps are the critical ones. This makes sense because the controllers must deal with a large amount of data received and control actions to be sent. In contrast, from a mission point of view, the pumps guarantee the system's operation. Pumps act directly on the dynamics of the water. 

\begin{figure}[!ht]
    \centering
    \includegraphics[width=0.9\linewidth]{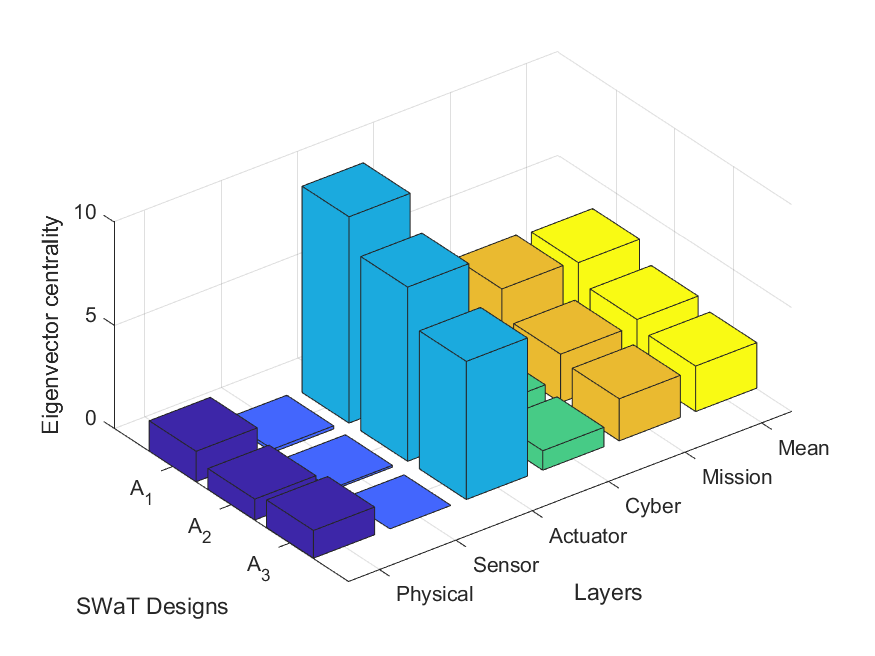}
    \caption{Graphical results of \gls*{swat} designs eigenvector centrality assessment presented in Table~\ref{tab_eig_results}.}
    \label{fig:eig_3d_results}
\end{figure}

Table~\ref{tab_eig_results} presents the normalized resilience assessment results obtained with the Eigenvector centrality metric. Fig.~\ref{fig:eig_3d_results} presents a graphical view of these results. These results present the estimated resilience of each layer for the three designs of \gls*{swat}. $A_1$ is the original design built according to the available technical details provided in the iTrust documentation \cite{swattestbed2021itrust}. $A_2$ is similar to $A_1$ with additional sensors, e.g., pump temperature and pump rotation speed. Thus, this architecture has higher monitorability capacities. $A_3$ is similar to $A_2$, with auxiliary pumps controlled by redundant controllers. These additional steerability capacities implies also an increase of the monitorability capacities to cover the new elements included.

\begin{table}[!t]
\footnotesize
\setlength{\tabcolsep}{4pt}
\renewcommand{\arraystretch}{1.5}
\begin{center}
\caption{Eigenvector centrality assessment of \gls*{swat} designs. Results are presented according to the multilayered model described in Section~\ref{sec:multi_layer}. Green cells represent the lowest eigenvector centrality values. In opposition, red cells show the highest eigenvector centrality values.}
\begin{tabular}{ c|c|c|c|c|c|c| }
\cline{2-7}
&
\textbf{Physical} &
\textbf{Sensor} &
\textbf{Actuator} &
\textbf{Cyber} &
\textbf{Mission} &
\textbf{MEAN} \\
\hline
\multicolumn{1}{ |l| }{\gls*{swat} $A_1$} & 
   \cellcolor[HTML]{ffb1b1} $1.4747911$ &
   \cellcolor[HTML]{ffb1b1} $0.1534561$ &
   \cellcolor[HTML]{ffb1b1} $10$ &
   \cellcolor[HTML]{ffb1b1} $2.2824026$ &
   \cellcolor[HTML]{ffb1b1} $3.6581340$ &
   \cellcolor[HTML]{ffb1b1} $3.5137568$ \\
   \cline{1-7}
\multicolumn{1}{ |l| }{\gls*{swat} $A_2$} &
   \cellcolor[HTML]{b5ffb1} $1.0110036$ &
   \cellcolor[HTML]{fff3b1} $0.0913183$ &
   \cellcolor[HTML]{fff3b1} $8.4442910$ &
   \cellcolor[HTML]{fff3b1} $1.1495613$ &
   \cellcolor[HTML]{fff3b1} $2.3502676$ &
   \cellcolor[HTML]{fff3b1} $2.6092884$ \\
   \cline{1-7}
\multicolumn{1}{ |l| }{\gls*{swat} $A_3$} &
   \cellcolor[HTML]{fff3b1} $1.3430652$ &
   \cellcolor[HTML]{b5ffb1} $0$ &
   \cellcolor[HTML]{b5ffb1} $6.6950694$ &
   \cellcolor[HTML]{b5ffb1} $0.9416670$ &
   \cellcolor[HTML]{b5ffb1} $2.0406492$ &
   \cellcolor[HTML]{b5ffb1} $2.2040902$ \\
\hline

\end{tabular}

\label{tab_eig_results}
\end{center}
\end{table}

Our results show a decrease in eigenvector centrality with increased resilience across the three designs. As we add more components to $A_2$ and $A_3$, the relative importance of each node in the graph decreases. There are more paths to reach each node, and the system's functions can be ensured with additional components, implying that we gain resilience.

\section{Experimentation}
\label{sec:experimentation}

\noindent In this section, we present our experimental results with \gls*{cps} metrics.

\subsection{Experimental Results}

\noindent Our approach assumes that we face adversaries that can spy on cyberspace to acquire knowledge about the system. In fact, critical infrastructures are becoming more and more connected due to increased competitiveness, leading to a race to digitization. Most complex system architectures components are connected to a network, send data to controllers, or communicate with other elements. An adversary able to spy on cyberspace can find critical points in the same way as we did. To achieve this goal, an adversary can use available online resources to build its model of \gls*{swat}. We conducted this experimentation using the \gls*{llm} Knowledge Graph Builder provided by Neo4j. This tool allows us to train our model to build a knowledge graph of \gls*{swat}. We can train our model by using documents, e.g., word or pdf files, images, unstructured text, and online resources from Wikipedia, YouTube, or other websites.

We used the following references, including a technical report provided by iTrust \cite{swattestbed2021itrust}, a conference paper by Mathur~\etal \cite{mathur2016swat}, and two iTrust videos available online \cite{swatintroyoutube,swatattackyoutube} to train our model. Then, by generating the graph using Openai gpt 4o as \gls*{llm} model, we obtain a knowledge graph representing the \gls*{swat} system. The graph obtained contains structural, logical, and organizational information.

Using the Neo4j Bloom exploration tool, the adversary can download its own model, explore the graph using queries, and investigate how the system works.

In the interface, a Neo4j knowledge graph chat is also available. Using the data extracted and analyzed from the uploaded documents makes it possible to ask questions to the ChatBot. For example, we asked: \textit{Can you describe in detail how SWaT works, how the components are connected to each other, and what type of data they exchange with each other?} We obtained a very detailed answer describing the sub-processes, how data are stored for log analysis, the network protocols, the topology, and also data exchanged between \glspl*{plc} and sensors or actuators, but also between \glspl*{plc} and the \gls*{scada} system. The answer provided by ChatBot is available in this repository~\cite{github_cyberresilienceanalytics}.

This first part of our experimentation shows how the adversary can generate its own \gls*{swat} model to find critical points. The Neo4j graph builder is handy for mapping nodes and entities in specific categories. These categories can be components, sensors, interfaces, or computers. An adversary attempting to attack a critical system can use a search engine such as Shodan to find vulnerabilities associated with specific components.

Shodan works like a search engine with a query language, which can be used for testing purposes. For example, the following command triggers a search for devices connected to the Internet with a default password set as ``1234'' connected from the city of Taipei in Taiwan.

\definecolor{codegreen}{rgb}{0,0.6,0}
\definecolor{codegray}{rgb}{0.5,0.5,0.5}
\definecolor{codepurple}{rgb}{0.58,0,0.82}
\definecolor{backcolour}{rgb}{0.95,0.95,0.92}

\lstdefinestyle{mystyle}{
    backgroundcolor=\color{backcolour},   
    commentstyle=\color{codegreen},
    keywordstyle=\color{magenta},
    numberstyle=\tiny\color{codegray},
    stringstyle=\color{codepurple},
    basicstyle=\ttfamily\footnotesize,
    breakatwhitespace=false,         
    breaklines=true,                 
    captionpos=b,                    
    keepspaces=true,                 
    numbers=left,                    
    numbersep=5pt,                  
    showspaces=false,                
    showstringspaces=false,
    showtabs=false,                  
    tabsize=2
}

\lstset{style=mystyle}

\begin{lstlisting}
''password 1234'' city:taipei
\end{lstlisting}

Shodan allows you to find the IP address with the searched vulnerability. Then, with a Putty connection, an adversary can attempt to connect to the corresponding computer. 

Let us now consider a concrete scenario. The case of the Oldsmar water treatment station in Florida in 2021, reported by Greenberg, illustrates the impact that an adversary can generate after gaining access to the system \cite{greenberg2021hacker,hardlessonsoldsmar}. The attack can be analyzed with the MITRE ATT\&CK \cite{mitre} catalog to find the associated sequence.

\begin{figure}[!ht]
    \centering
    \includegraphics[width=0.8\linewidth]{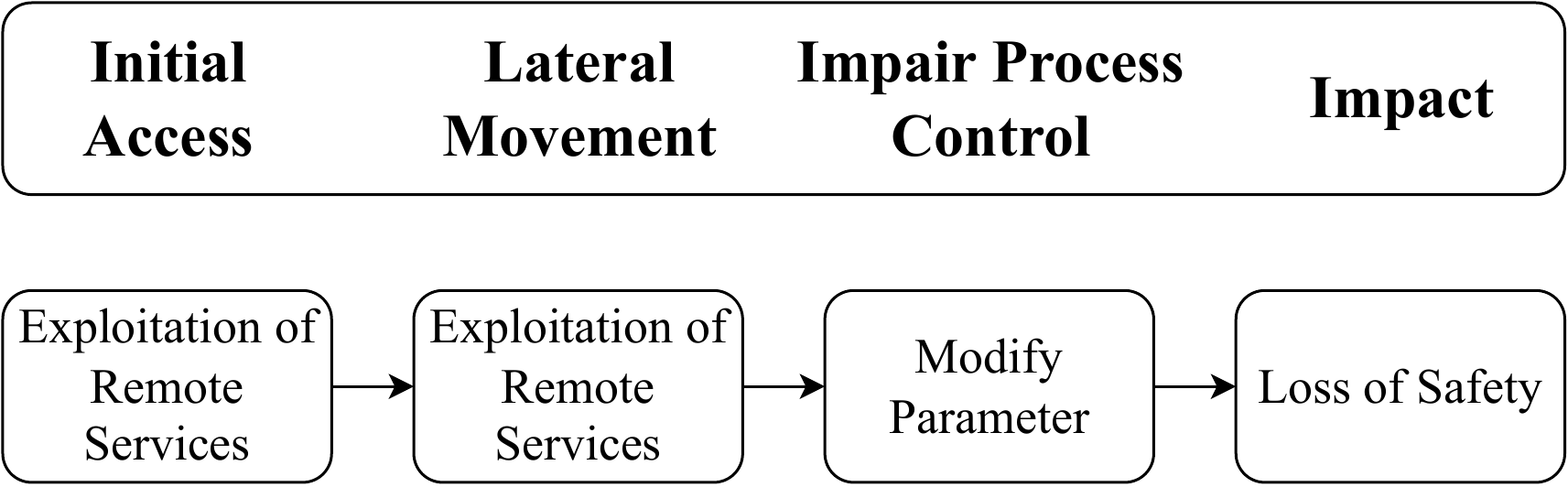}
    \caption{Tactics that could be used by \glspl*{apt} to perpetrate a similar attack to the Oldsmar one against a water treatment facility.}
    \label{fig:mitre_simplified}
\end{figure}

We conducted an analysis applied to the case of the Oldsmar cyber attack \cite{hardlessonsoldsmar}. We created two layers in the MITRE ATT\&CK navigator. The first is used to select \glspl*{ttp} related to \glspl*{ics} used by the fifteen threat groups identified in MITRE ATT\&CK v16. The second is used to identify \glspl*{ttp} employed to perpetrate an attack similar to the Oldsmar water treatment facility in 2021. In 
\ref{app:mitre}, Fig.~\ref{fig:mitre_swat} shows the complete MITRE ATT\&CK view. A simplified view presenting the common tactics extracted that real adversaries could be able to perpetrate against a water treatment facility is presented in Fig.~\ref{fig:mitre_simplified}.

This analysis illustrates that well-known \glspl*{apt} can reproduce a similar attack perpetrated against \glspl*{ics}.

According to the available reports regarding the description of the Oldsmar attack, we learn that the adversary used a human error for which the TeamViewer associated to the system interface was unintentionally exposed to the Internet. This could occur when a server is exposed without a password requirement.

With the following command, Shodan can find the Print Server Web exposed without any password required for authentication:

\begin{lstlisting}
PRINT_SERVER WEB +200 -401 -NeedPassword
\end{lstlisting}

For this demonstration, we used specific queries dedicated to finding devices associated with water treatment facilities. Consider Canada as a location for conducting our investigation. The associated query is:

\begin{lstlisting}
water treatment country:CA
\end{lstlisting}

We obtained five results. Two of them are associated with the same architecture located in Moosomin. Searching on Google, we learn that a new water treatment plant in Moosomin will open up in May 2025 \cite{moosomin2025}.

By investigating the results provided by Shodan, we learn that a router provided by the manufacturer MikroTik is used in this infrastructure. We also learn that a \gls*{fftx} technology is associated with one of the two IP addresses given by Shodan. We learn that ports 161 and 2222 that support TCP/UDP are open on the two IPs.

In a PowerShell window, the following commands confirm that port 2222 is open.

\begin{lstlisting}
Test-NetConnection @IP -Port 161
Test-NetConnection @IP -Port 2222
\end{lstlisting}

According to CVE reports, there are vulnerabilities associated with port 2222. CVE-2007-0655 stipulates that the MicroWorld Agent Service allows remote or local adversaries to gain privileges and execute arbitrary commands by connecting to port 2222~\cite{cveport2222}.

This experimentation presents how the adversary could act to perpetrate an attack similar to the Oldsmar one. We remind you that resilience considers the fact that we are facing powerful adversaries and that attacks are possible. Thus, they are bound to happen. The objective of resilience is to establish barriers that make attacks very difficult to carry out.

We can identify critical points in an architecture and an adversary. Once identified, we can protect them and deploy appropriate countermeasures to avoid cascading effects.

\subsection{Analysis of Results}
\label{sec:analysis_results}

\begin{figure}[!ht]
\centering
\includegraphics[width=\textwidth]{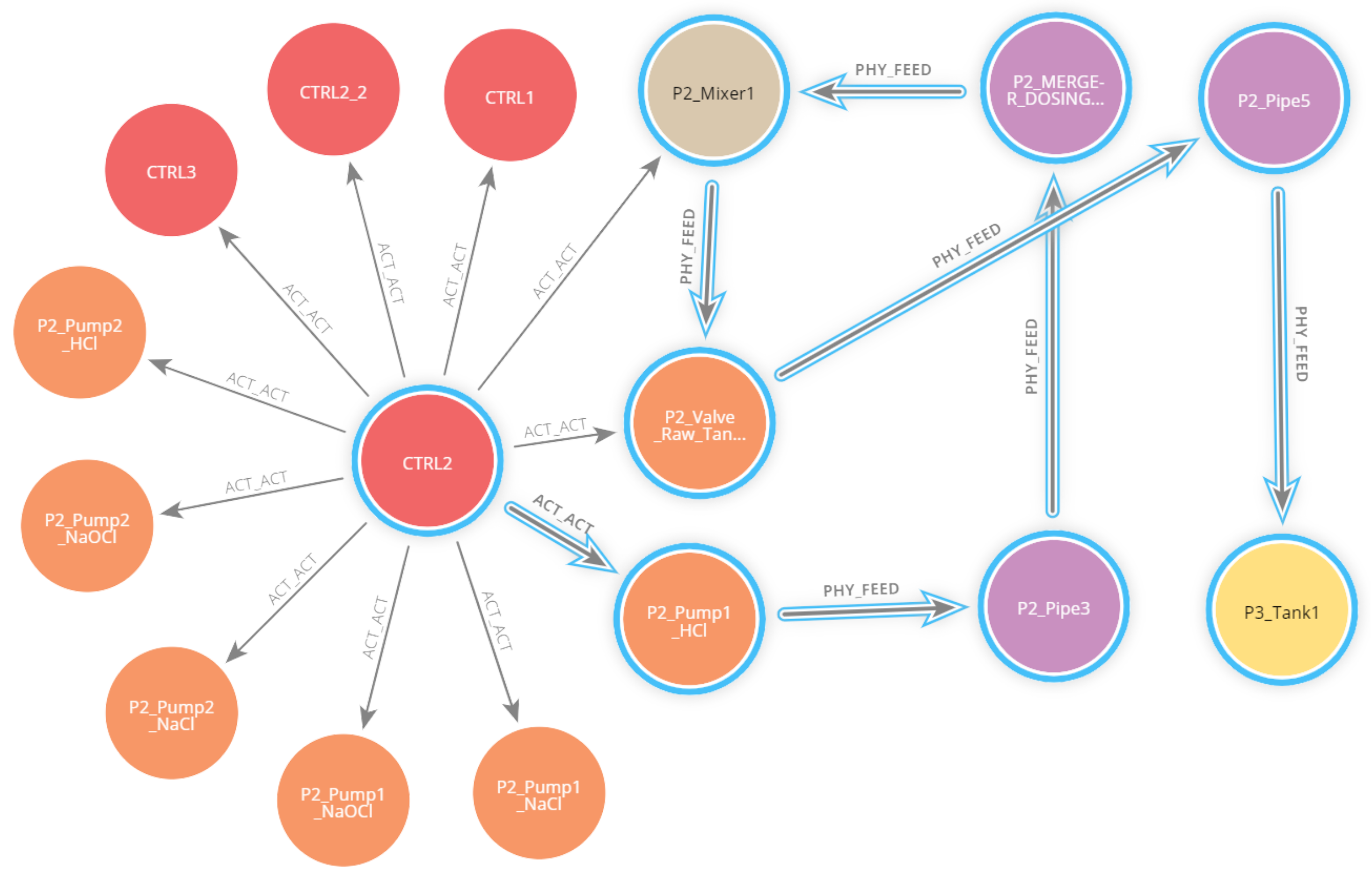}
\caption{Sub-graph of \gls*{swat} original architecture $A_1$.}
\label{fig:uc_a1}
\end{figure}

\noindent An adversary attempting to attack a system can use malicious techniques such as those presented in this section. In the case of the Oldsmar cyber attack, the adversary tried to poison the water by adding a quantity of sodium hydroxide that was one hundred times higher than the usual amount.  This could have been possible if the \gls*{swat} design $A_1$ had been attacked. Fig.~\ref{fig:uc_a1} shows a subgraph of $A_1$. Consider a similar scenario to Oldsmar in which the adversary compromises the controller of the chemical dosing stage. The blue path shows that the attack impact is propagated to a pump that adds hydrochloric acid ($HCl$) to the water. Once the product has been added, we see the water being driven to a mixer through a pipe network before reaching the third stage (yellow node). The attack cannot be absorbed without enough monitorability to detect the attack and controllability to react.

The architecture $A_2$ is more monitorable than $A_1$, with a set of supplementary sensors. Fig.~\ref{fig:uc_a2} shows a subgraph in the same chemical dosing process in $A_2$. Two supplementary sensors, i.e., temperature sensor and rotation speed sensor, are attached to the pump. These sensors can detect abnormal behavior.

\begin{figure}[!ht]
    \centering
    \includegraphics[width=\textwidth]{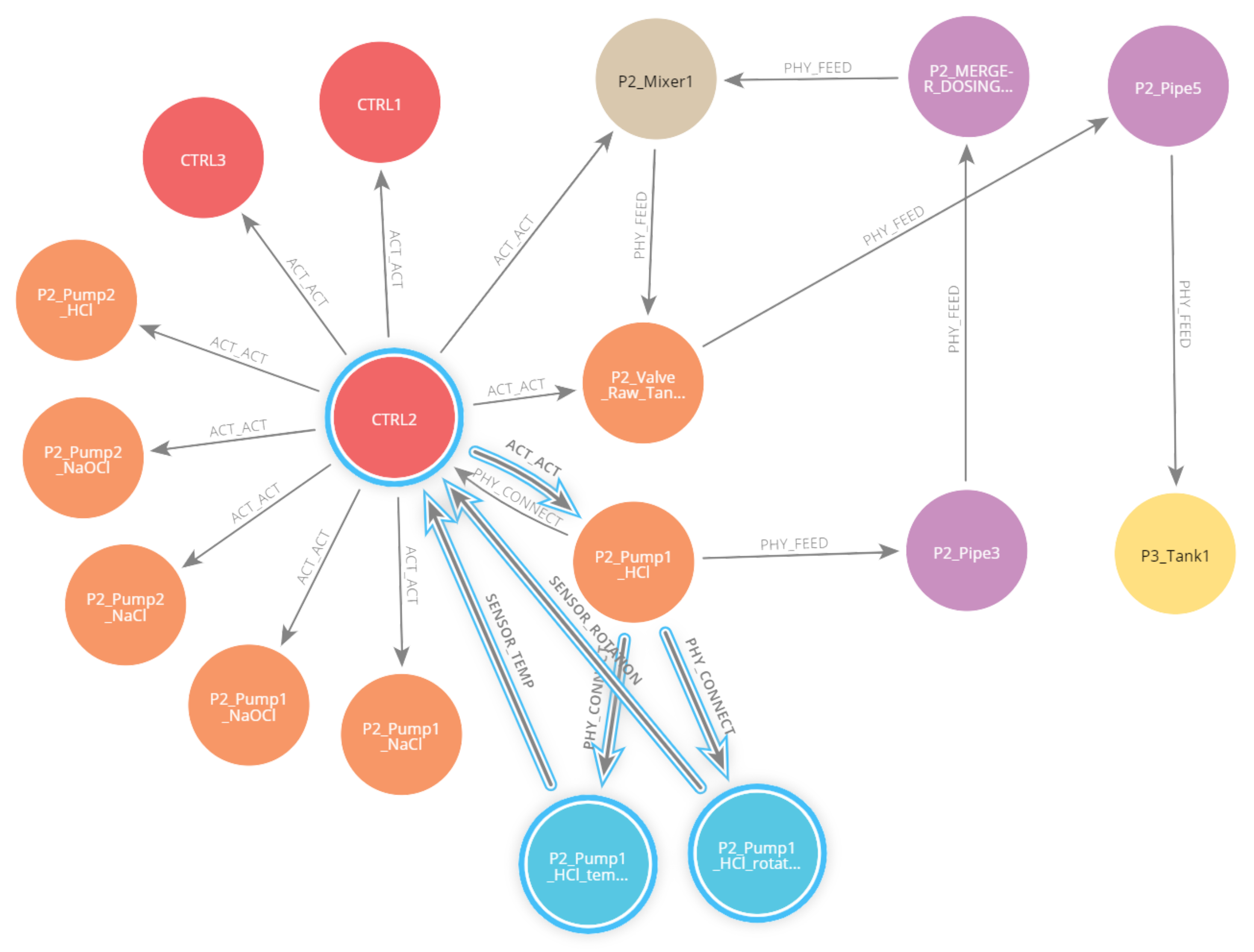}
    \caption{Sub-graph of \gls*{swat} architecture with more monitorability $A_2$.}
    \label{fig:uc_a2}
\end{figure}

\begin{figure}[!ht]
    \centering
    \includegraphics[width=\textwidth]{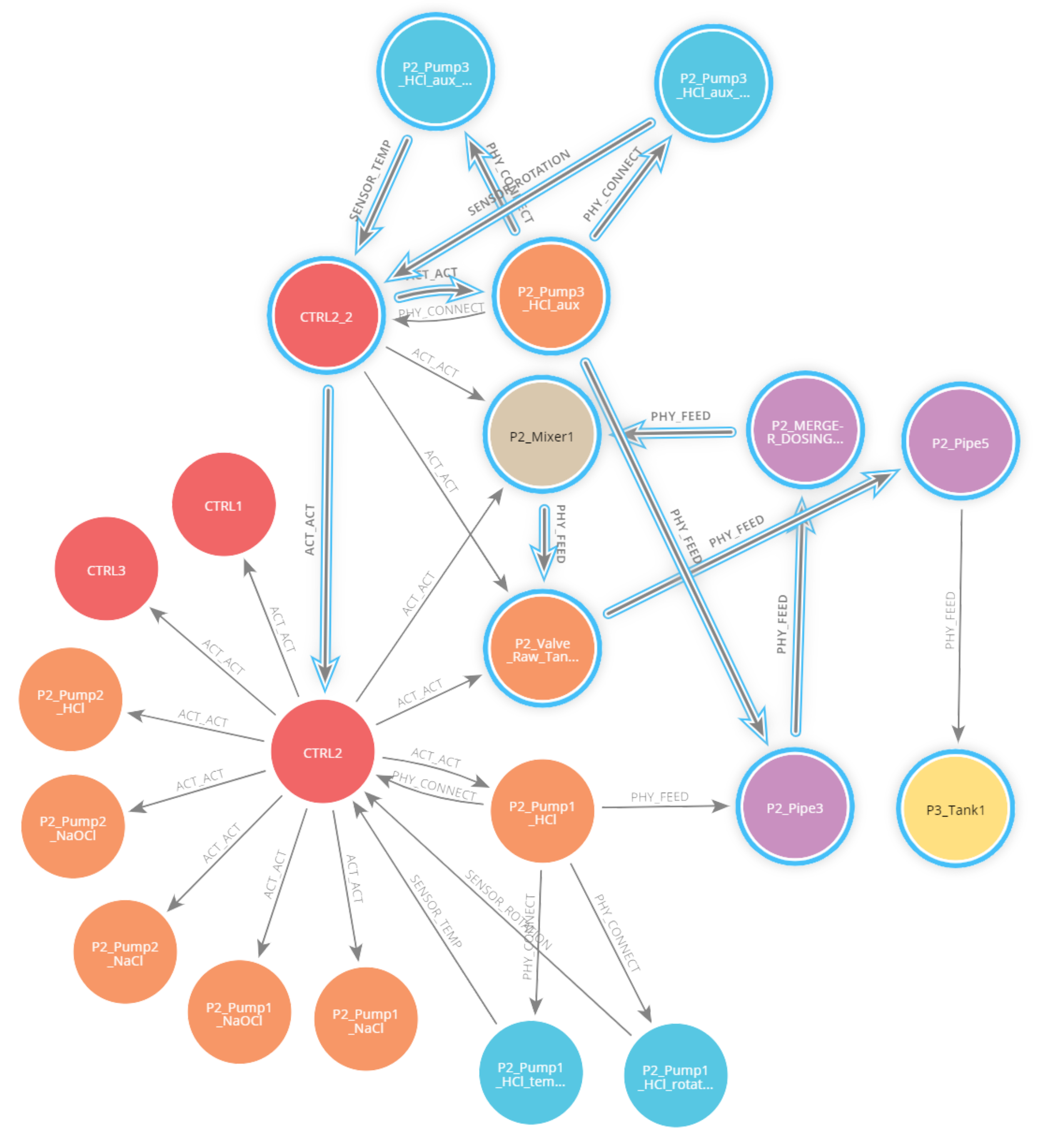}
    \caption{Sub-graph of \gls*{swat} architecture with more monitorability and steerability $A_3$.}
    \label{fig:uc_a3}
\end{figure}

Finally, the architecture $A_3$ is more monitorable and steerable than $A_2$, with supplementary auxiliary actuators, including pumps and controllers. Fig.~\ref{fig:uc_a3} shows a subgraph in the same chemical dosing process in $A_2$. The auxiliary controller can shut down the compromised one and take control to start an auxiliary pump.

\begin{figure}[!ht]
\centering
\includegraphics[width=0.9\textwidth]{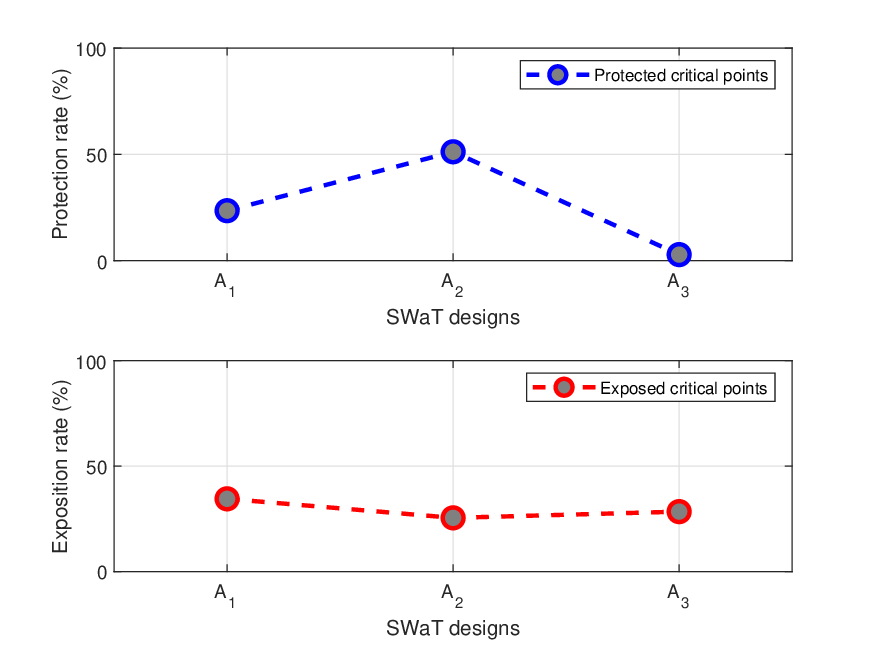}
\caption{Protection and exposition rate of \gls*{swat} three designs critical points.}
\label{fig:protection_exposition}
\end{figure}

$A_3$ has the most desirable resilience potential to detect and absorb cyber-attacks. However, ensuring the protection of critical points is also essential. Fig.~\ref{fig:protection_exposition} is a graphical representation of the protection and exposition rates related to the \gls*{swat} critical points. The protection rate, in blue, was calculated for each design as the ratio between the number of protected critical points and the total number of essential nodes identified in each design. The exposition rate, in red, has been computed as the ratio between the number of nonprotected critical points and the total number of components in each design.

Having a resilient architecture may imply exposing more critical points to cyber adversaries. Indeed, the architecture is more complex, and even if the relative importance of each node decreases due to adding more components, critical points still exist that can generate cascading effects. Thus, there is a delicate balance to find in the conception phase to build resilient designs with few critical points exposed to cyber adversaries.

\subsection{Discussion}

\noindent Our results show that $A_3$ is the most resilient architecture according to the eigenvector centrality results. The proposed methodology also allows for identifying critical points in each layer of our model. The proposed experiment shows that $A_3$ has a sufficient resilience potential to detect the attack and react to bring the system back to a stable state. Regarding remediation, our approach does not consider the time required to absorb the attack and bounce back to a stable state. Different techniques based on learning methods exist, such as those described by Cai~\etal \cite{cai2024survey} for considering temporal knowledge graphs. In our case, for quantifying resilience and identifying critical points, we used static knowledge graphs to emphasize the structural aspects of a design that can bring resilience capacities.

We must also highlight that quantification methods based on attacks' impact can also be established based on multilayered models. Indeed, the relationships between components, i.e., data exchanged, required information, or control actions, can help establish attack impact quantification methods based on losses.

\section{Conclusion}
\label{sec:conclusion}

\noindent In this paper, we presented a methodology based on multilayered modeling of \glspl*{cps}. We compared the results of three graph analytics metrics on three resilient designs of \gls*{swat}. We identified the eigenvector centrality metric as the most relevant for quantifying resilience and identifying critical points. Our analysis identifies the critical points of each architecture that an adversary can target to generate an attack with cascading effects. We presented a methodology that a defender could use to anticipate adversarial knowledge gain about the architecture design, including truth discovery for values manipulated by an adversary. We used the Neo4j LLM Graph Builder tool and trained our model with online content, including YouTube videos and publications related to the \gls*{swat}. This analysis shows that an adversary manipulating the graph locally with the Neo4j Desktop version can use specific plugins to identify the most critical points. Based on this very same knowledge, the defender can anticipate the malicious actions of the adversary to protect those identified critical points.

We foresee the following perspectives for quantifying and improving the resilience of complex systems in future work. Firstly, analyzing components from different manufacturers to find vulnerabilities that can open backdoors to cyber adversaries. Secondly, considering attack impact quantification strategies can help improve resilience with an additional layer, such as a dependency layer, including dependencies between components. 

\bibliographystyle{elsarticle-num} 
\bibliography{Biblio}

\newpage

\appendix

\section{MITRE ATT\&CK and Water Facility Scenario}
\label{app:mitre}

\begin{figure}[!ht]
\centering
\includegraphics[width=0.99\textwidth]{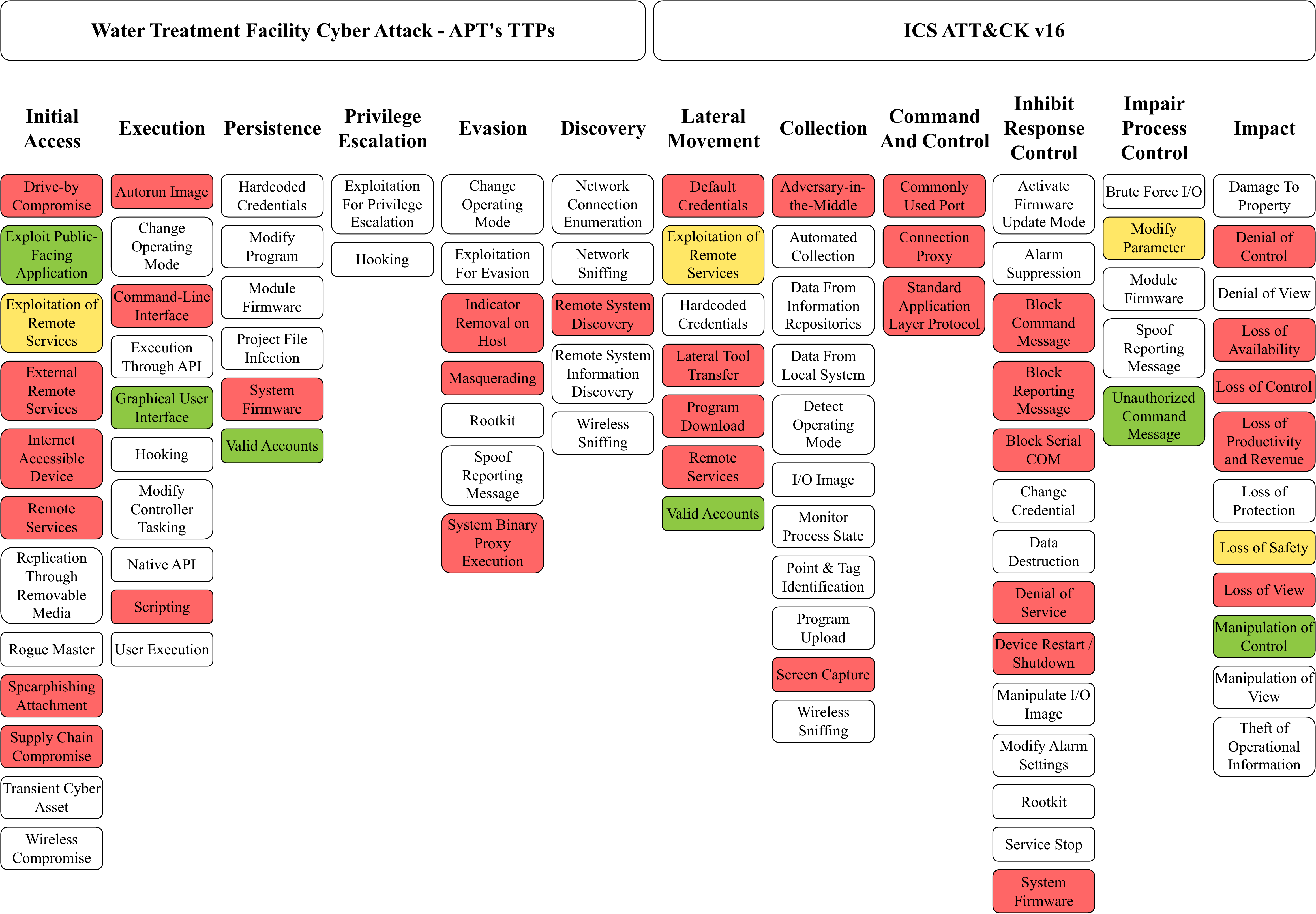}
\caption{MITRE ATT\&CK guidelines~\cite{mitre} applied to a water facility cyber attack. Red and green tactics are respectively related to the first and second layers, namely \glspl*{ttp} related to \glspl*{ics} used by the fifteen threat groups identified in MITRE ATT\&CK v16 and \glspl*{ttp} employed to perpetrate an attack similar to the one of the Oldsmar water treatment facility in 2021. Yellow tactics are the common ones between the two layers.}
\label{fig:mitre_swat}
\end{figure}

\end{document}